\documentclass{aa}  

\usepackage{graphicx}
\usepackage{txfonts}
\usepackage{amsmath}
\usepackage{booktabs}
\usepackage{xcolor}
\usepackage{ulem}
\newcommand{\nicer}{{\it NICER}}
\newcommand{\chandra}{{\it Chandra}}
\newcommand{\maxi}{{\it MAXI}/GSC}

\begin{document}

   \title{ALMA/\nicer\/ observations of GRS 1915$+$105 indicate a return to a hard state}

   \titlerunning{ALMA/\nicer\/ observations of GRS 1915$+$105}

   \author{K. I. I. Koljonen\thanks{\email{karri.koljonen@utu.fi}}
          \and
          T. Hovatta
          }

   \institute{The Finnish Centre for Astronomy with ESO,
              University of Turku\\
            Aalto University Mets\"ahovi Radio Observatory, Mets\"ahovintie 114, FI-02540 Kylm\"al\"a, Finland \\
             }

   \date{Received ; accepted }

 
  \abstract
   {\object{GRS 1915+105} is a transient black hole X-ray binary consistently emitting 10--100\% of the Eddington luminosity in the X-ray band over the last three decades until mid-2018 when the source luminosity suddenly decreased by an order of magnitude. This phase was followed by a change to a state with even lower average X-ray fluxes never seen before during the outburst but presenting renewed flaring activity at different wavelengths, albeit with mean fluxes still in decline.}
   {\object{GRS 1915+105} has the longest orbital period known among low-mass X-ray binaries, the largest accretion disk size, and therefore the largest mass supply for accretion. The high inclination of the disk allows the study of geometrical effects of the accretion flow such as changes in the height-to-radius ratio or the effect of accretion disk winds on the intrinsic emission that is expected during the outburst decay. In addition, the transient jet is expected to change to a compact, self-absorbed, steady jet.}
   {We conducted two full polarization \textit{Atacama Large Millimeter Array} observations to study the jet properties during the outburst decay by analyzing the spectral, polarization, and intra-epoch variability for both observation epochs. In addition, we analyzed almost daily \textit{Neutron Star Interior Composition Explorer} pointing observations, modeling X-ray power spectral densities, spectral energy distributions, and light curves with a physically motivated model to follow the changing accretion disk properties throughout the outburst decay and relating them to the jet emission.}
   {We show that the X-ray and millimeter (mm) spectral, timing, and polarization properties are consistent with those of a typical decaying X-ray binary outburst and that \object{GRS 1915+105} has descended into the low-luminosity hard X-ray state. The jet emission in the mm is consistent with a compact, steady jet with $\sim$1\% linear polarization, and the magnetic field is likely aligned with the jet position angle. Relating the mm emission to the X-ray emission reveals that the source has changed from a higher radio/X-ray correlation index to a lower one; $L_{\mathrm{radio}} \propto L_{X}^{0.6}$ .}
   {}


   \maketitle
%

\section{Introduction}

\object{GRS 1915+105} is a transient, black hole X-ray binary (XRB) that started its decades-long outburst in 1992 \citep{castrotirado94}. During the outburst, it was one of the brightest XRBs, consistently emitting 10--100\% of the Eddington luminosity in the X-ray band. However, in mid-2018, the source luminosity decreased suddenly by an order of magnitude and has diminished since. \object{GRS 1915+105} hosts a K-type giant star \citep{greiner01} with a mass of $\sim$0.5 solar masses as a donor \citep{steeghs13}, firmly establishing it as a low-mass XRB. It has the longest orbital period known among low-mass XRBs (33.9 days; \citealt{steeghs13}) and therefore the largest accretion disk size and mass supply for accretion. 

\object{GRS 1915+105} was the first Galactic source known to exhibit apparent superluminal motion of the jet components \citep{mirabel94}. The source presented both transient and steady jet phases during its outburst. The transient jet phases occurred approximately once every year and consisted of bright events at infrared and radio wavelengths  \citep{pooley97,kleinwolt02}. In between the transient jet phases, the steady jet phases were periods of prolonged hard X-ray emission \citep{foster96,dhawan00,fuchs03} and a low radio flux density of $\sim$10 mJy \citep{ogley00} indicating emission from a compact self-absorbed jet. The transient jet phase in XRBs often coincides with a time when the X-ray spectrum changes rapidly. This spectral evolution is taken to arise from a physical change in the accretion disk structure where the optically thin and geometrically thick disk is replaced by an optically thick and geometrically thin disk due to increased mass accretion rate \citep{gallo03}. During this time, strong radio flares are observed, and bright, optically thin synchrotron ``blobs'' or internal shocks can be seen emanating from the central source in radio interferometric images \citep{fender99}. 

Due to the strong radio emission and trackable jet components, the jet and system parameters of \object{GRS 1915+105} are known relatively well. The jet component velocities range between 0.65$c$ and 0.98$c$, jet inclination is 60-70 deg to the line of sight, the mass of the black hole is 10-14 solar masses, and the distance is 7-10 kpc \citep{fender99,reid14}. Due to the high inclination, we are looking at the accretion disk relatively edge-on. This allows us to potentially study the evolution of the accretion flow geometry, such as changes in the height-to-radius ratio of the accretion flow or the effect of accretion disk winds on the intrinsic emission \citep[e.g.][]{lee02,miller16,neilsen18}.

In July 2018, \object{GRS 1915+105} entered an unusually extended low-flux X-ray phase followed by a change to a state with even lower average X-ray fluxes never seen before during the outburst but presenting renewed flaring activity at different wavelengths, most notably in the radio \citep{koljonen19b,trushkin19,motta19,motta21}. Detailed X-ray observations suggest that diminished and obscured accretion might be feeding the variable jets \citep{koljonen20,miller20,neilsen20,motta21,balakrishnan21}. The obscuring matter is likely located close to the X-ray source either ejected from the accretion disk by a disk wind or originating from a puffed-up or warped accretion flow.

In this paper, we present two full polarization \textit{Atacama Large Millimeter Array} (ALMA) and nearly daily \textit{Neutron Star Interior Composition Explorer} (\nicer) monitoring observations of \object{GRS 1915+105} that were taken during the outburst decay before the source entered to the obscured state. The observations and data reduction processes are described in Section 2. In Section 3, we go through the detailed timing and spectral analyses of both \nicer\/ and ALMA data sets. We show that the X-ray and millimeter (mm) spectral and timing properties of \object{GRS 1915+105} show consistent behavior of a decaying XRB outburst indicating that \object{GRS 1915+105} has descended to a low-luminosity hard state. The outburst decay of \object{GRS 1915+105} offers us a detailed view of the physical processes in the accretion flow and the jet leading from the outburst towards quiescence and to the anomalous obscured state. In Section 4, we discuss the accretion disk and jet properties during the different phases of the outburst decay, and speculate on accretion disk properties in a scenario where the outburst has ended. Finally, we conclude in Section 5. 

\section{Observations}

\subsection{ALMA}\label{almaobs}

We conducted two band 3 (90--105 GHz) full polarization ALMA observations in Cycle 6 on November 1, 2018, and March 21, 2019, during the outburst decay of \object{GRS 1915+105}. Both epochs lasted approximately two hours, consisting of several alternating scans of the science source ($\sim10$ min), polarization calibrator ($\sim$2 min), and phase calibrator sources ($\sim$0.5 min). The polarization calibrator source in both epochs was J1924$-$2914, and the phase calibrator source was J1922$+$1530 in the first epoch and J1905$+$0952 in the second epoch.

We performed the data reduction with the \textit{Common Astronomy Software Applications} (CASA) package version 5.4.0-70 by running the calibration script provided with the data. We flagged the first ten channels in the second spectral window due to bad D-terms for both epochs. In addition, antennas DA63 and DA41 for the first and second epochs, respectively, were flagged for all spectral windows due to outlier values in gain amplitude. Also, for the second epoch, we flagged antennas DA42 and DA48 for the first spectral window due to outlier values in the X- and Y-polarized phase difference. The resulting images show that the source is a point source within the beam ($\sim$0.6\arcsec\/ and $\sim$2.5\arcsec\/ in the first and second epochs, respectively) without any observed structure. 

Flux densities were compared with both imaging methods and a delta function was fitted to the uv-plane using \textsc{uvmultifit 3.0.0} \citep{martividal14}; we find that they agree with each other. We estimated the Stokes Q and U polarized fluxes averaging over all spectral windows and all scans. To conduct spectral and timing studies, we estimated Stokes I fluxes for each spectral window individually averaging in time over the whole observations, and for each scan length by averaging over all spectral windows. We made the same analysis for the science, polarization, and phase calibrator sources. 

To verify that the variability in the target on scan-length integrations is not due to instrumental effects, we did the following tests. We examined the gain amplitudes from the phase calibration to verify that any changes in the amplitudes are not correlated with the flux density variations. We also examined the water vapor radiometer data of each antenna to detect any correlated changes in atmospheric conditions and the target flux density, but none were present. The mean precipitable water vapor content during the first and second epochs was 1.3\,mm and 0.9\,mm, respectively. Combined with the more extended configuration during the first epoch, we expect a larger scatter in the flux density due to changing atmospheric conditions during the first epoch. This is indeed seen in our third check where we divided the array into two individual sub-arrays (DV and DA antennas separately) and repeated the flux density estimation. During the first epoch, the flux density estimates from the sub-arrays vary up to 0.2\,mJy, and during the second epoch they vary by less than 0.1\,mJy.

\subsection{NICER} \label{nicer}

We downloaded all \nicer\/ data from the High Energy Astrophysics Science Archive Research Center (HEASARC) using a time range of MJD 58238--58634 (May 2018 -- May 2019) beginning from the outburst decay phase and ending in the descent to the obscured phase. We disregarded observations with exposure times of less than 500 seconds. This resulted in 87 pointings, shown in Table A1.

We reduced the observations using \textsc{nicerdas} version 6a with parameters \textsc{nicercal\_filtexpr="EVENT\_FLAGS=bx1x000",} which removes all undershoot, overshoot, forced trigger, and fast-only events, and \textsc{cor\_range="4-"} to remove high particle radiation intervals associated with the Earth's auroral zones. We selected PI energy channels between 30 and 1200 (0.3--12 keV). We extracted the X-ray spectra using \textsc{xselect v2.4g}. We used the photon redistribution matrix (RMF) and the on-axis average ancillary response file (ARF) for all 52 detectors combined from HEASARC's Calibration Database (CALDB). For the background we use a public background file available in HEASARC (\textsc{nixtiback20190807.pi}).

For the timing analysis, we calculated the power spectral densities (PSDs) directly from the cleaned event files using a binning of 2$^{-7}$ s (or 2$^{-4}$ s in case of the data in the obscured phase), an energy band 1--10 keV, and a segment length of 16 s (or 256 s in case of the data in the obscured phase). We normalize all the PSD to rms variability. All the segments were further averaged over the whole pointing and binned geometrically by a factor of 1.05 before importing them to the Interactive Spectral Interpretation System (\textsc{isis}; \citealt{houck00}) for model fitting.  

For the spectral analysis, we binned the data adaptively in the ranges 0.3-1.5 keV to S/N=5-10, 1.5-5.0 keV to S/N=20-70, 5.0-8.0 keV to S/N=15-50, and 8.0-12.0 keV to S/N=10-30 depending on the flux and exposure time of the pointing to ensure similar statistics for each spectrum. We performed spectral fitting using \textsc{isis}. We estimated the errors on the parameter values and fluxes through Monte Carlo analysis. For those spectra with 1-10 keV flux densities below 0.3$\times$10$^{-8}$ erg s$^{-1}$ cm$^{-2}$, an additional unabsorbed power-law component was needed to nicely fit the low-energy data below 1 keV that could arise from imperfect background subtraction. However, the normalization of this component is very small, containing only 0.003-0.006\% of the total flux and affecting only the soft X-rays below 1 keV with a fairly steep spectral slope of $\Gamma\sim$2.5.

\section{Results}

\subsection{X-ray monitoring} \label{monitoring}

 \begin{figure}
 \centering
 \includegraphics[width=\linewidth]{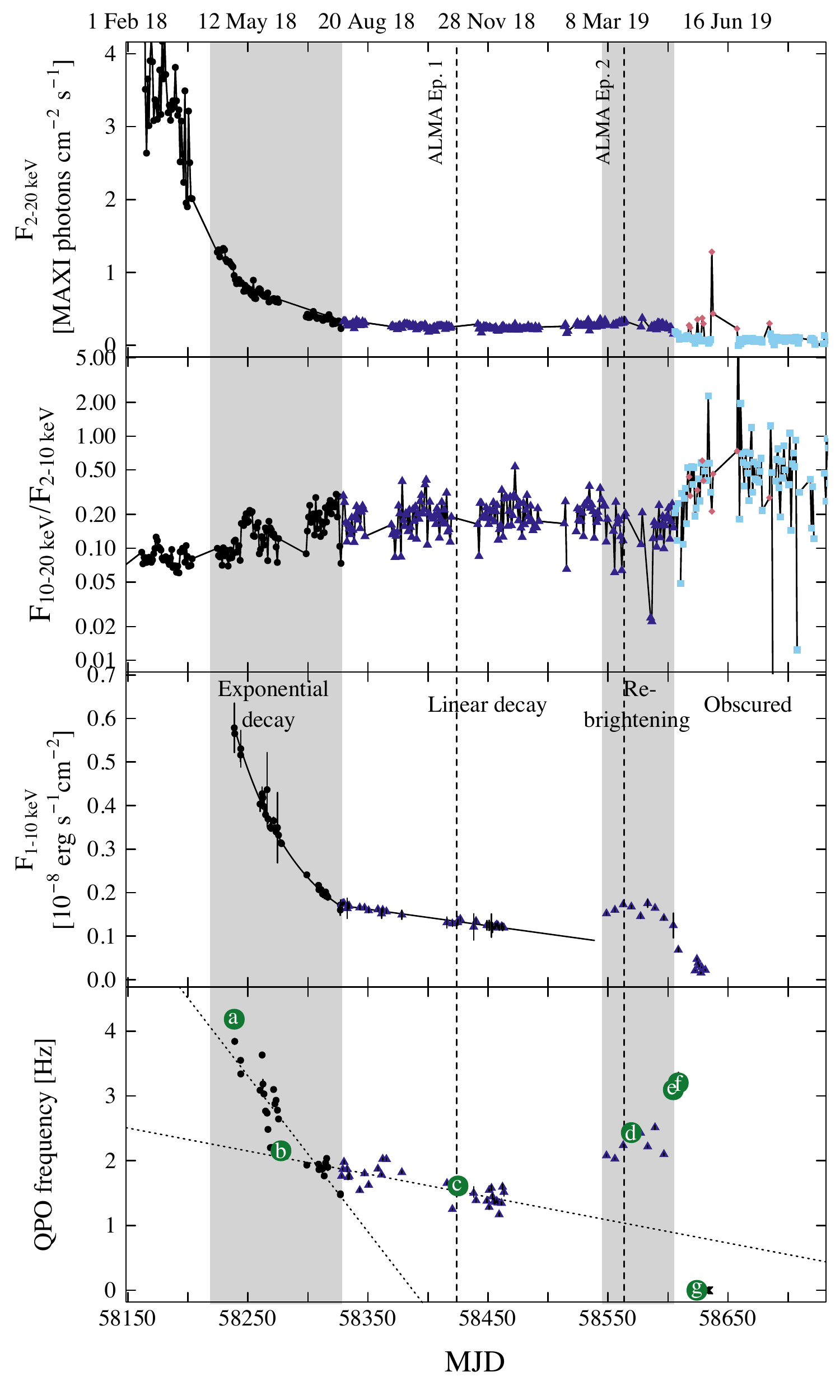}
 \caption{\textit{Top}: \maxi\/ 2--20 keV daily light curve of \object{GRS 1915+105} since February 2018 with ALMA observations marked as dashed vertical lines. The data points are colored and marked according to the hardness-intensity diagram shown in Fig. \ref{hid}. \textit{Middle/top}: \maxi\/ 10--20 keV / 2-10 keV daily hardness ratio. \textit{Middle/bottom}: \nicer\/ 1--10 keV light curve with flux densities estimated from spectral fitting and the light curve decay fitted with an exponential and a linear decay models (solid line). The different decay phases are indicated and shown as an alternating shading scheme. \textit{Bottom}: QPO frequencies as determined from modeling the \nicer\/ PSD. The exponential and linear decay phases show different rates of decay for the QPO frequency (shown as dotted lines). The seven lettered observations correspond to example PSD and spectra shown in Figs. \ref{psd}, \ref{psd_g}, and \ref{nicerspec}.} 
 \label{asm}
\end{figure} 

 \begin{figure}
 \centering
 \includegraphics[width=\linewidth]{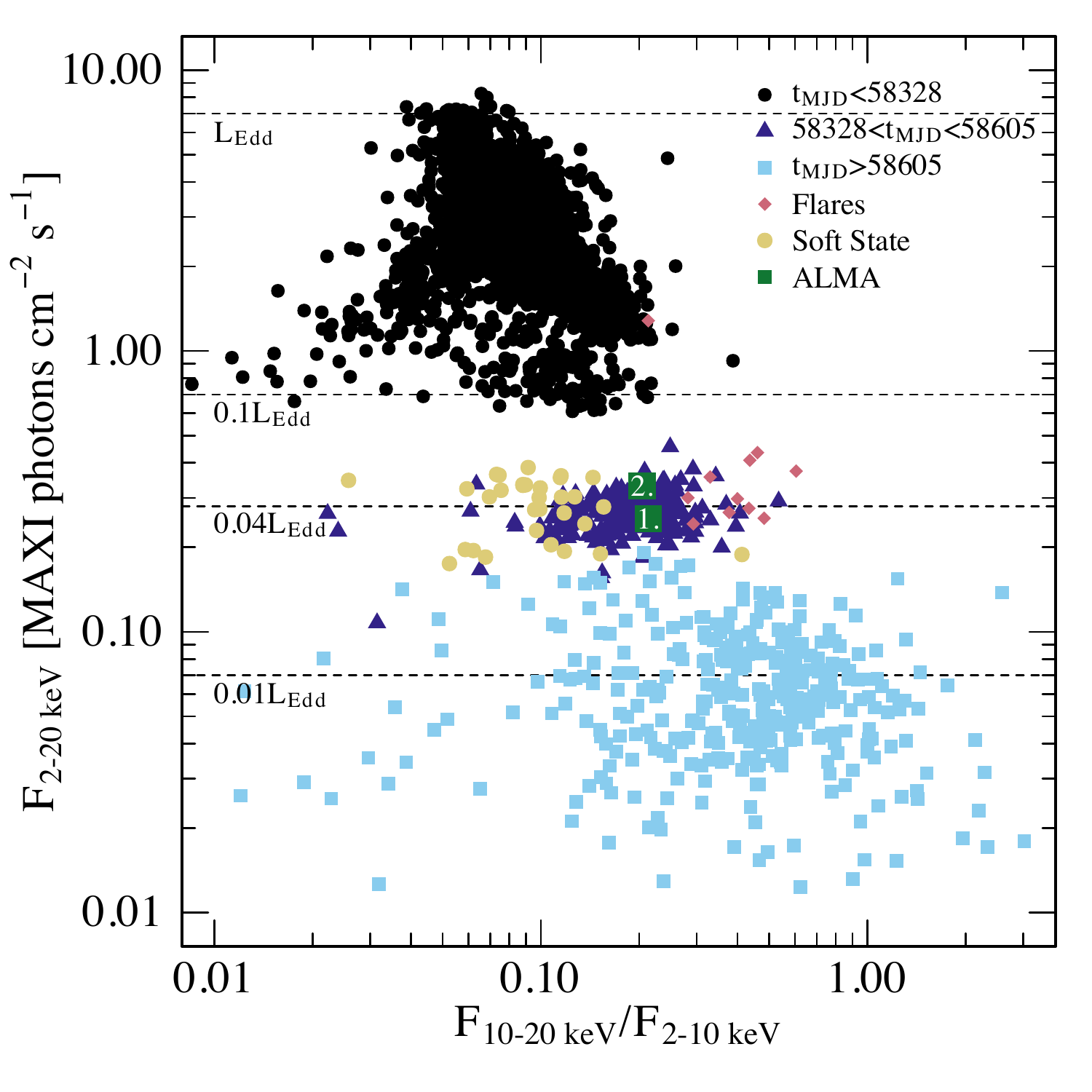}
 \caption{\maxi\/ hardness-intensity diagram of \object{GRS 1915+105} from daily monitoring observations since August 2009. The blue data points (both dark triangles and light blue squares) indicate the low-luminosity state since August 2018 with increased spectral hardness. The light blue squares correspond to the obscured decay phase with even lower flux densities and harder spectra with occasional, strong X-ray flares (red diamonds) and highly variable radio emission, in addition to a softer state in August-September 2020 with the flux densities back to the level of the low-luminosity state (yellow dots). The numbered green boxes correspond to the ALMA epochs that took place during the linear and rebrightening decay phases of the outburst decay.}
 \label{hid}
\end{figure} 

The ALMA observations coincided with a stable low-luminosity X-ray state where the observed luminosity is much lower and X-ray hardness higher than the usual outburst values. This state was superseded by an even lower luminosity state with a higher X-ray hardness ratio that is most likely an effect of obscuration \citep{koljonen20,miller20,neilsen20,motta21,balakrishnan21}. Figure \ref{asm} shows the \textit{Monitor of All-sky X-ray Image}/Gas Slit Camera (\maxi) 2--20 keV count rate, \maxi\/ hardness ratio, \nicer\/ 1--10 keV flux density, and quasi-periodic oscillation (QPO) frequencies as measured from the PSD during the decay states: black circles denoting the exponentially decaying outburst, blue triangles denoting a linearly decaying phase, and light blue squares denoting the obscured state (with sporadic flares denoted as red diamonds). In between the linear decay phase and the obscured phase, \object{GRS 1915+105} exhibited an increase in flux density and QPO frequency departing from the linear decay trend, which we denote as a rebrightening phase. 

Figure \ref{hid} shows the hardness-intensity diagram from the \maxi\/ daily observations of \object{GRS 1915+105} since Aug 2009. The three decay states separate easily from each other with a change in the countrate and the X-ray hardness. Assuming that \object{GRS 1915+105} is accreting at the Eddington limit for the highest count rates ($\sim$7 photons cm$^{-2}$ s$^{-1}$) ---a reasonable argument based on both observations (e.g., \citealt{done04}; but taking into account the effect of the smaller distance estimate in \citealt{reid14}) and simulations \citep[e.g.,][]{truss04}--- the outburst phase corresponds to luminosities $0.1 < L/L_{\mathrm{Edd}} < 1.0$, the linear decay phase to $\sim$0.04 L$_{\mathrm{Edd}}$, and the obscured phase to $\sim$0.01 L$_{\mathrm{Edd}}$ (which agrees with values obtained by \citealt{koljonen20} and \citealt{miller20}). During the obscured phase, there were prominent flares with X-ray luminosities reaching $\sim$0.04--0.18 L$_{\mathrm{Edd}}$, and more recently, in August-September 2020, a softer state with X-ray luminosities around $\sim$0.04 L$_{\mathrm{Edd}}$. Assuming a black hole mass of 12 M$_{\odot}$ \citep{reid14} the Eddington luminosity is L$_{\mathrm{Edd}} = 1.5 \times 10^{39}$ erg s$^{-1}$.  

\subsection{X-ray spectral and timing properties during the outburst decay phases}

In this section, we describe the X-ray spectral and timing properties during the outburst decay in detail. As described above, we divide the data in four separate phases with distinct behavior in the X-ray light curve as well as in X-ray spectral and timing properties.

\subsubsection{Description of the spectral model}

We fit the individual pointing spectra from the \nicer\/ dataset with a model consisting of an absorbed power law and reflection model components (\textsc{relxill}; \citealt{garcia14,dauser14}) modified by Gaussian emission and/or absorption line components when necessary. In addition, for spectra with 1-10 keV fluxes below 0.3$\times$10$^{-8}$ erg s$^{-1}$ cm$^{-2}$ we added an unabsorbed power-law model component to take into account the effect of imperfect background subtraction as mentioned in Section \ref{nicer}. 

For the photoelectric absorption component, we chose \textsc{vphabs,} which allow the abundances of the elements of the neutral absorber to be changed. \object{GRS 1915+105} is known to be surrounded by local, cold material likely in the form of an accretion disk wind that absorbs the X-ray emission in addition to the interstellar absorption \citep{lee02,martocchia06}. Following \citet{martocchia06}, we grouped elements that have either a small effect in the energy range in question or whose origin is likely to be the same into the following groups: (1) H, He, C, N, O, Ne, Na, (2) Mg, (3) Al, (4) Si, (5) S, (6) Cl, Ar, Ca, and (7) Cr, Co, Ni, and Fe. Depending on the quality of the spectrum, we grouped the abundances to an even smaller number of groups (the smallest division consisting of two groups with one for lighter elements than aluminum and one for heavier elements). In the fitting process, we found that the iron abundance tended towards unphysically low values, and therefore we decided to fix it to the solar value. In addition, we added emission and/or absorption lines of neutral and ionized iron (Fe I K$\alpha$ at 6.40 keV, Fe XXV K$\alpha$ complex at 6.7 keV, Fe XXV K$\beta$ at 7.80 keV, Fe XXVI Ly$\alpha$ at 6.97 keV, and Fe XXVI Ly$\beta$ at 8.27 keV), sulfur (2.3 keV), argon (3.0 keV), and calcium (3.7 keV) when necessary, mainly for the better quality spectra. Thus, the total model can be described as: \textsc{vphabs} $\times$ (\textsc{relxill} + \textsc{lines}) + \textsc{powerlaw}.

In the fitting procedure, we first fitted the spectra with the largest exposure times to pinpoint the values of some of the \textsc{relxill} parameters that were then kept fixed when modeling the lower quality spectra. These included the inner radius of the accretion disk R$_{\mathrm{in}}$, spin of the black hole $a$, and the reflection factor R$_{\mathrm{f}}$ of \textsc{relxill}. However, we found that the spin is not constrained, and the inner radius presents large values indicating that the X-ray spectra are not sensitive to relativistic effects. Therefore, we decided to fix the spin to zero. We further fixed the inclination to 70 degrees \citep{reid14,mirabel94}, and iron abundance to the solar value. Other parameters were left as default values and fixed, except the power-law photon index ($\Gamma$), ionization parameter (log $\xi$), and model normalization, which were allowed to vary for all spectral fits.

 \begin{figure*}
 \centering
 \includegraphics[width=\linewidth]{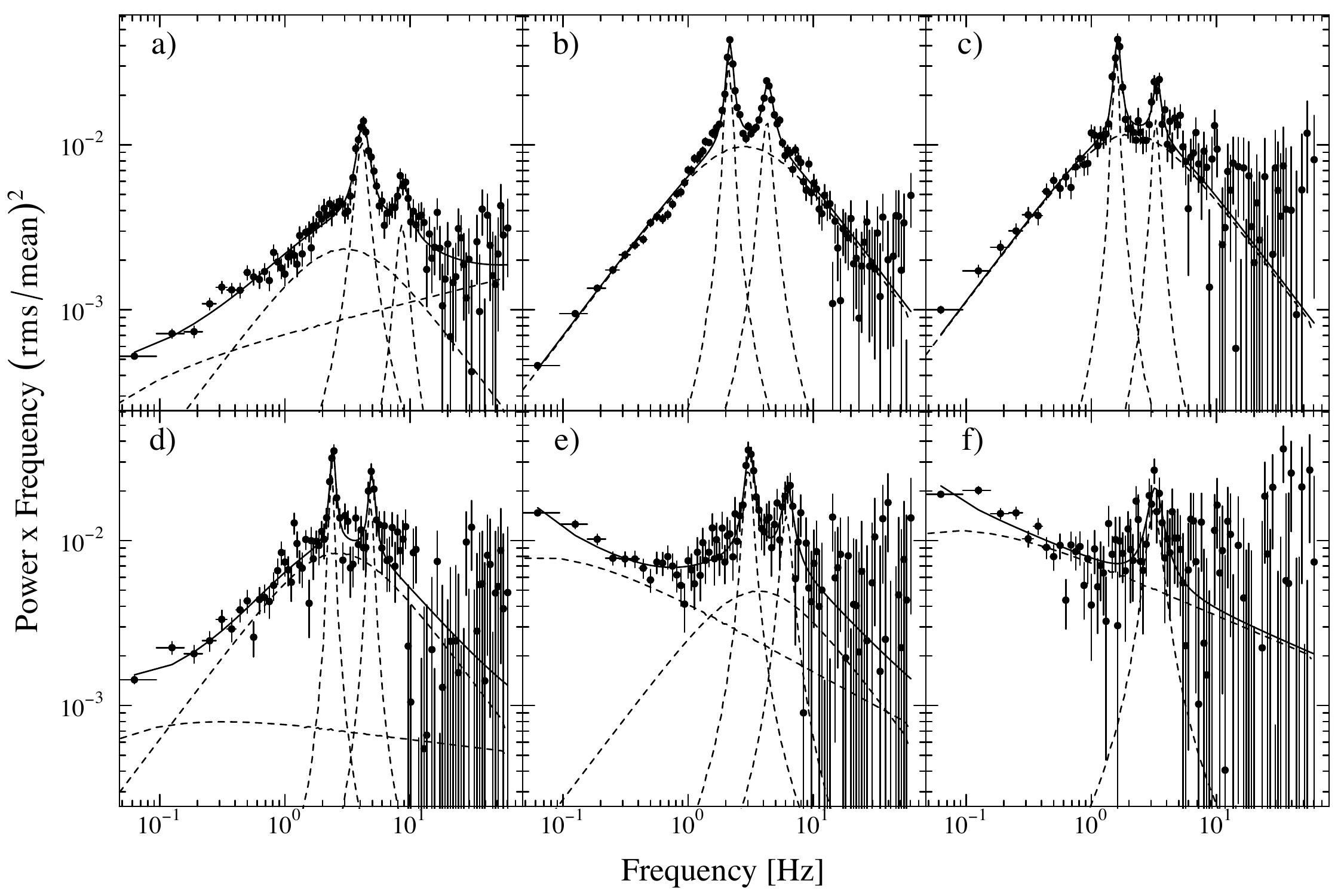}
 \caption{PSD from six different epochs labeled in Fig. \ref{asm} (\textit{bottom panel}). The PSDs are fitted with a model consisting of a zero-frequency Lorentzian modeling the broadband noise, one or two narrow Lorentzians modeling the QPO (fundamental and first upper harmonic components), an additional power-law component needed for PSD in panels \textit{a}, \textit{d}, \textit{e}, and \textit{f}, and Poisson noise as a constant power component (removed from the PSDs shown here).}
 \label{psd}
\end{figure*} 

\subsubsection{Description of the PSD model} \label{psd_model}

We fit all the PSDs with up to three Lorentzians: a Lorentzian centered at zero frequency for the band-limited noise and two Lorentzians for the QPO and its upper harmonic. In addition to the Lorentzian components, a power-law component is needed to model the higher luminosity PSD during the high-luminosity part of the exponential decay. We also model the Poisson noise present in the PSD as a constant power component.

\subsubsection{The exponential and linear decay phases}

In Fig. \ref{asm}, both the \nicer\/ 1-10 keV flux density and the QPO frequencies show similar evolution with a transition from a steeper (exponential) decay to a more shallow (linear) decay at $\sim$ MJD 58330. We take this change as the approximate transition time of the two decay phases. The \nicer\/ PSDs during the exponential and linear decay phases suggest a typical hard state PSD similar to what is observed in plateau and radio-quiet states with a band-limited noise component and a type-C QPO. The QPO frequency is tightly correlated with the X-ray luminosity decreasing approximately linearly from 4 Hz to 2 Hz during the exponential decay phase and further to $\sim$1.5 Hz during the linear decay phase and inversely correlated with the X-ray hardness ratio, both of which are typical for type-C QPOs \citep{casella05}.

We fit all the PSDs with the model described in Section \ref{psd_model} (examples are given in Fig. \ref{psd}, \textit{panels b--c}). The total rms of the Lorentzians correlates with the X-ray hardness ratio starting at $\sim$20\% at the beginning of the exponential decay phase and reaching $\sim$35\% in the linear decay phase. The rms evolution is mainly due to increasing rms of the zero-centered Lorentzian while the rms of the QPO and harmonic are on average 7$\pm$1\% and 6$\pm$1\%, respectively. The coherence parameter of the QPO component is always Q$>$5 indicating peaked noise. In addition to the Lorentzian components, a power-law component is needed to model the higher luminosity PSD during the high-luminosity part of the exponential decay (\textit{panel a} in Fig. \ref{psd}). It shows decreasing rms variability during the decay and presents a power-law index of $\Gamma\sim1.0$. All model parameters for each PSD are tabulated in Table A1.

 \begin{figure}
 \centering
 \includegraphics[width=\linewidth]{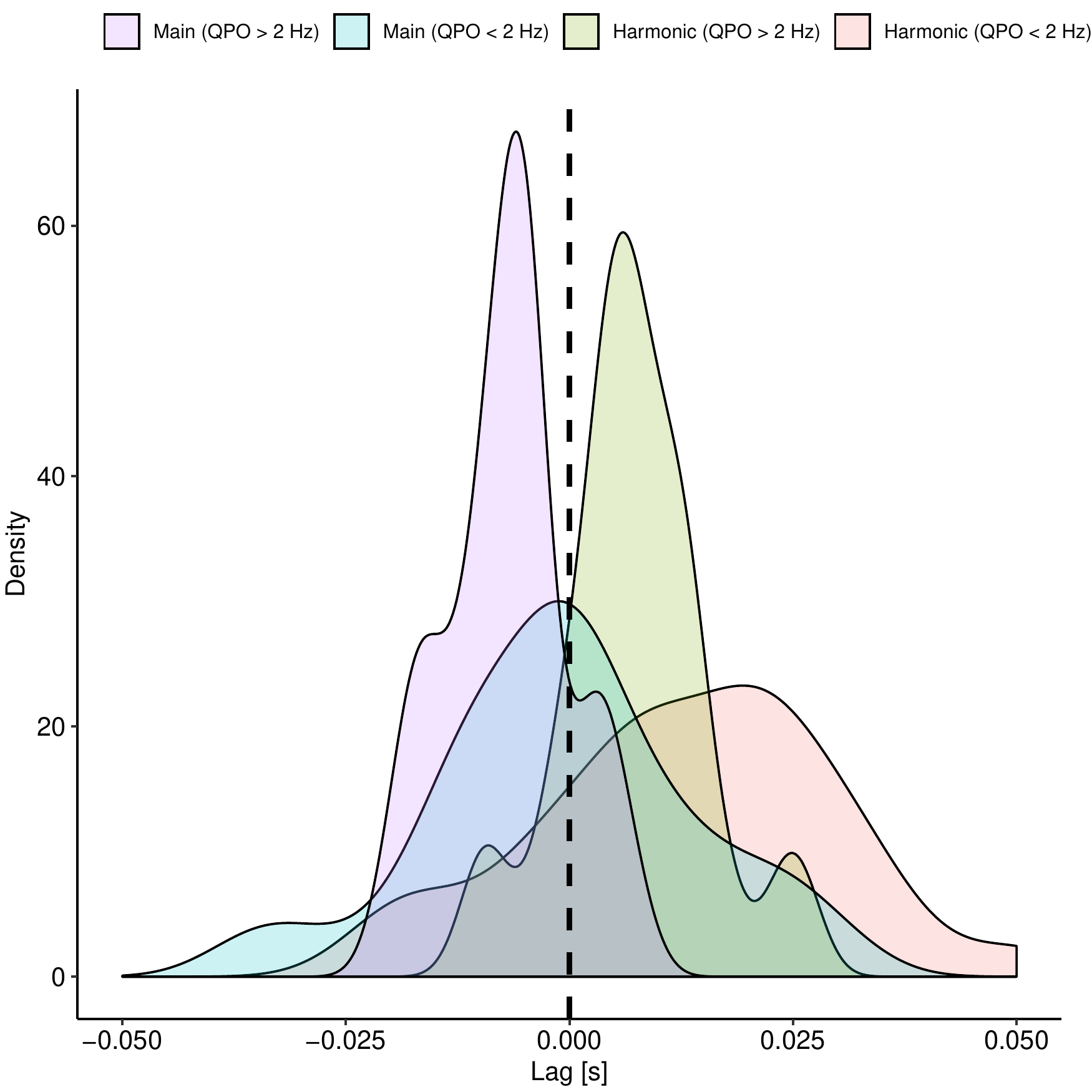}
 \caption{Lag distributions of the QPO frequencies between 2-4 keV and 5-10 keV bands. The main QPO frequency and the first upper harmonic frequency display, on average, soft and hard lags, respectively, when the main QPO frequency is above 2 Hz. When the QPO frequency is below 2 Hz, the average lag at the main QPO frequency vanishes, and the first upper harmonic QPO frequency shows hard lags.}
 \label{lag}
\end{figure} 

We also checked whether the QPO frequencies presented any time lags between a soft (2--4 keV) and a hard X-ray (5--10 keV) band (see Fig. \ref{lag}). On average, the main QPO frequency shows soft lags (soft photons lagging hard photons) and the upper harmonic hard lags (hard photons lagging soft photons) of about 10 msec when the QPO frequency is above 2 Hz. When the QPO frequency descends below 2 Hz; that is, during the linear decay phase, the lag at the QPO frequency averages to zero lag, but the upper harmonic frequency shows hard lags. This behavior is consistent with earlier studies during the outburst phase \citep{reig00,qu10,pahari13,zhang20}, where the phase lag has a log-linear relationship with the QPO frequency and changes sign approximately at 2 Hz.   

 \begin{figure}
 \centering
 \includegraphics[width=\linewidth]{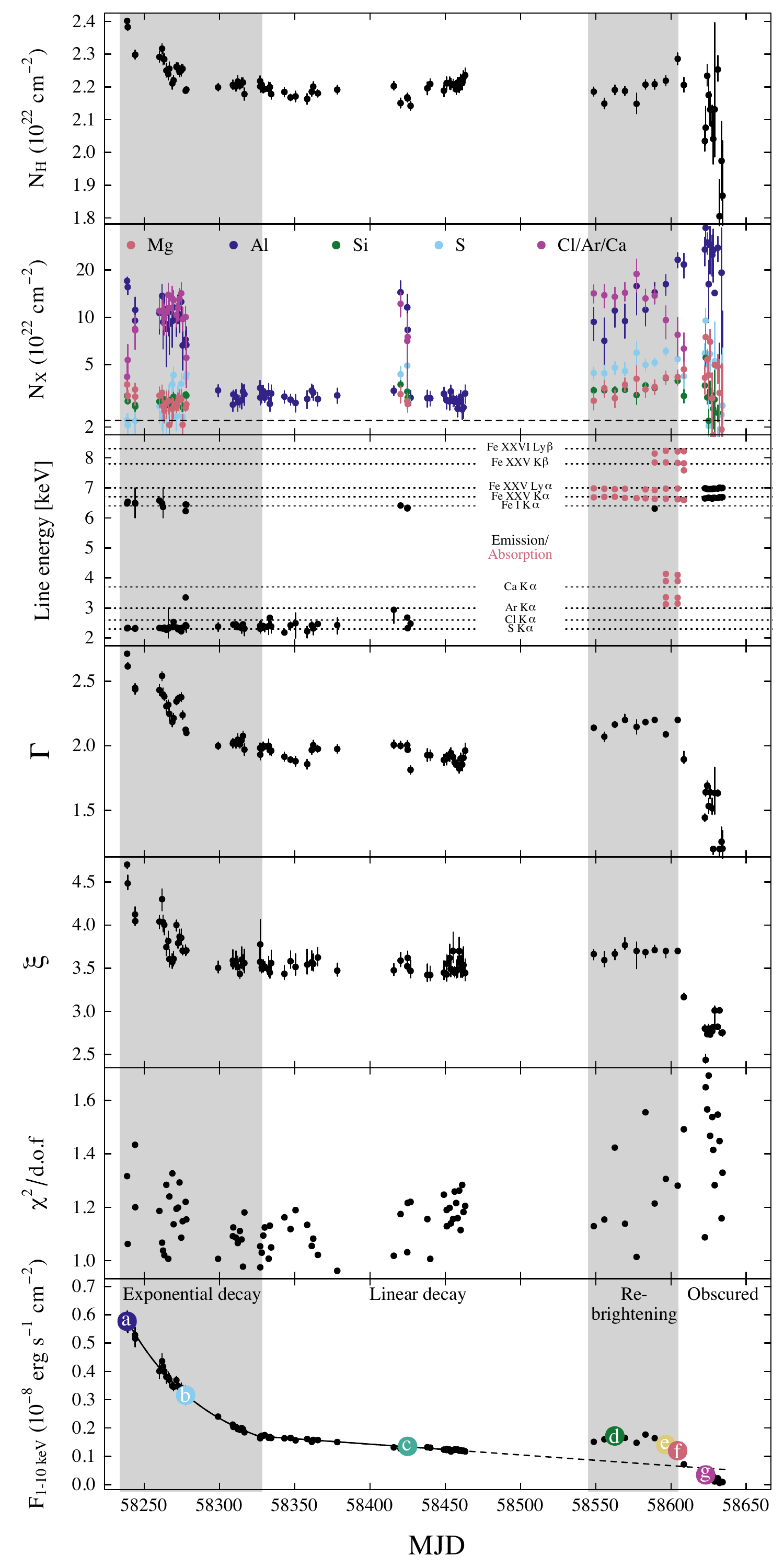}
 \caption{Spectral fit results of the \nicer\/ data. From top to bottom, the model parameter evolution is shown for the abundance of lighter elements (lighter than magnesium all fixed to the same value), heavier elements (Mg, Al, Si, S, and Cl/Ar/Ca displayed with different colors), line energies for the emission/absorption lines (displayed with different colors), the photon power law index ($\Gamma$), ionization parameter (log $\xi$), the corresponding fit quality ($\chi^{2}$/d.o.f), and the absorbed 1-10 keV flux density with the colored letters corresponding to spectra shown in Fig. \ref{nicerspec}. The alternate shading shows the times of the different decay phases.}
 \label{specfit}
\end{figure} 

  \begin{figure}
 \centering
 \includegraphics[width=\linewidth]{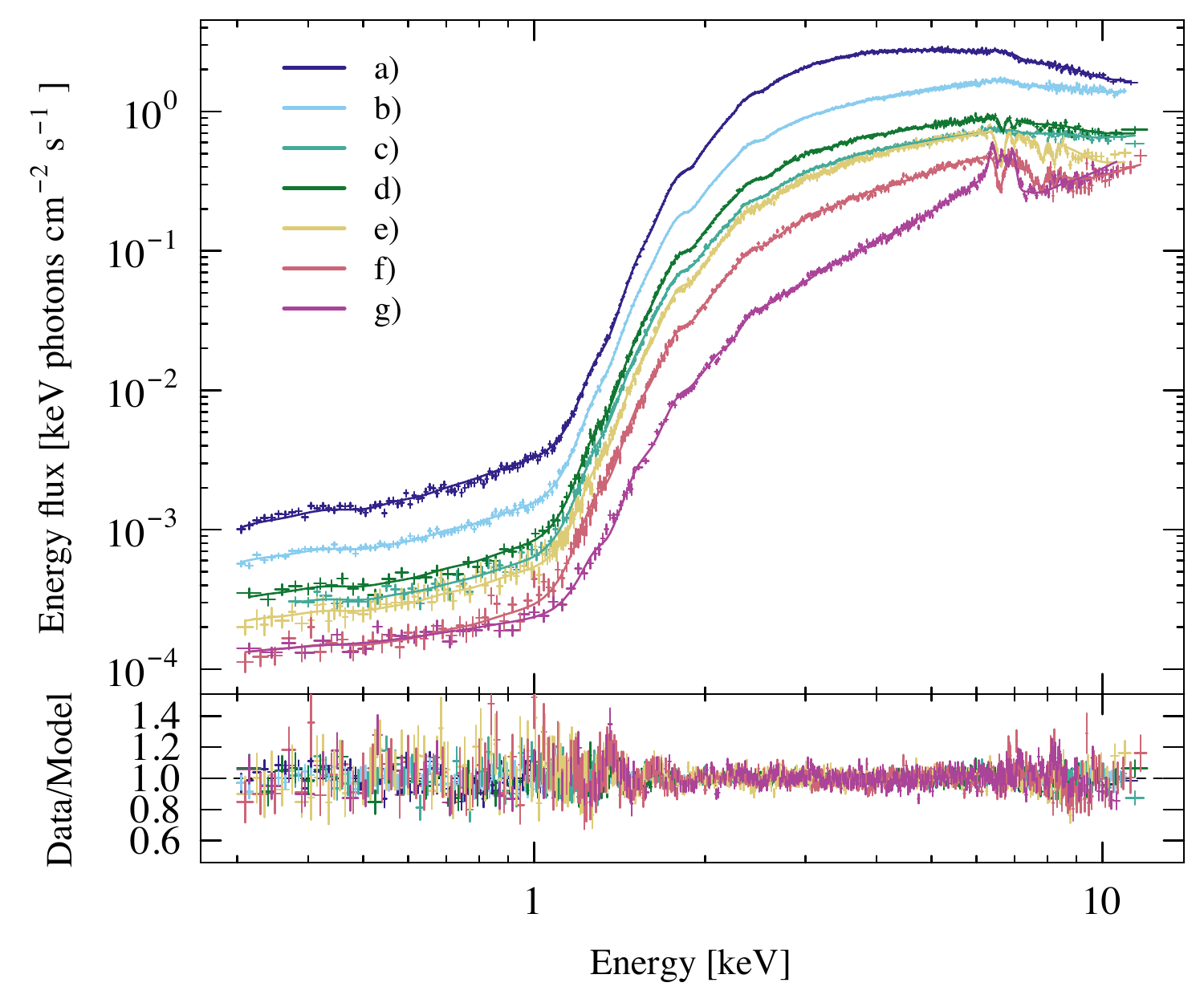}
 \caption{\nicer\/ spectra and best-fit models from the seven different epochs labeled in Fig. \ref{asm} (\textit{bottom panel}) and Fig. \ref{specfit} (\textit{bottom panel}). The coloring scheme is according to the colored letters shown in Fig. \ref{specfit} (\textit{bottom panel}).}
 \label{nicerspec}
\end{figure} 

During the exponential decay phase, the X-ray spectra show a slowly hardening and diminishing Comptonization component. The power-law photon index of the \textsc{relxill} component decreases from $\Gamma$ = 2.5 to $\Gamma$ = 2.0, ionization parameter from log $\xi$ = 4.5 to log $\xi$ = 3.5, and abundances of the absorber likely linked to the diminishing ionizing flux. The most prominent line features in the spectra are Fe I K$\alpha$ and S K$\alpha$. In the linear decay phase, the continuum model parameters settle to a more stable state with $\Gamma$=1.93$\pm$0.06 and log $\xi$ = 3.52$\pm$0.07. Typical values for the power-law index in an XRB hard state are $\Gamma$=1.5-2.0, while intermediate and soft states have $\Gamma>2$ \citep{remillard06}. The model parameter evolution is shown in Fig. \ref{specfit}, and the corresponding parameter values are provided in Table A2. In addition,  Fig. \ref{nicerspec} shows the spectra from the same \nicer\/ pointings as shown in Fig. {\ref{psd}}.

\subsubsection{The rebrightening phase}

After a gap in the monitoring data of \object{GRS 1915+105} (MJD 58460--58550, see Fig. \ref{asm}), the flux density and QPO frequency had departed from the decay profile as the source entered the rebrightening phase. During this phase, the flux density and the QPO frequency are not correlated although both present an increase of about a factor of two with the QPO frequency rising up to 3 Hz but vanishing completely during the flux drop leading to the obscured phase. 

In the X-ray PSD, a broad-band low-frequency component is visible in addition to a band-limited noise component and a type-C QPO (\textit{panels d--f} in Fig. \ref{psd}). As above, we model the additional component with a power-law model. The rms variability of the power-law component increases, and the power-law index presents higher values up to $\Gamma\sim1.5$. On the contrary, the rms of the zero-centered Lorentzian decreases slightly.    

In the X-ray spectra, the power-law photon index and the ionization parameter present elevated but steady values of $\Gamma$=2.14$\pm$0.06 and log $\xi$ = 3.73$\pm$0.05 as compared with the linear decay phase. The abundances show an increasing trend, high-ionization absorption lines appear in the spectra, and the emission lines disappear (Fig. \ref{specfit}, and Table A2). 

We further concentrated on the iron-line region towards the end of the rebrightening phase and in the obscured phase. Figure \ref{nicerabs} shows the evolution of the iron lines during the transition, and the corresponding line model parameters are presented in Table \ref{ironlines}. The rebrightening phase is dominated by the absorption lines with Fe XXV K$\alpha$ being the most dominant line with a resolved FWHM of 160-180 eV (corresponding to a velocity of 7000-8000 km/s) and equivalent width of about 100 eV. The line center is also gradually redshifted from 6.70 keV (corresponding to a zero redshift) to 6.59 keV (corresponding to a redshift of $\sim$4900 km/s). In addition, Fe XXV K$\beta$ and Fe XXVI Ly$\beta$ absorption lines are also visible and possibly Ni K$\alpha$ line in one pointing at MJD 58608.5, although the energy of the line is off by 0.1 keV. 

During the rebrightening phase, there is a clear evolution of the ratio of Fe XXVI Ly$\beta$ and Fe XXV K$\beta$ column densities (estimated according to \citet{lee02}; their Eq. 1) indicating a decreasing ionization factor from log $\xi\sim$ 3 to log $\xi\sim$ 2.6 \citep{kallman01}. These values are lower than what is obtained with the continuum fitting where log $\xi\sim$ 3.7 throughout the rebrightening phase.     

\subsubsection{The obscured phase}

 \begin{figure}
 \centering
 \includegraphics[width=\linewidth]{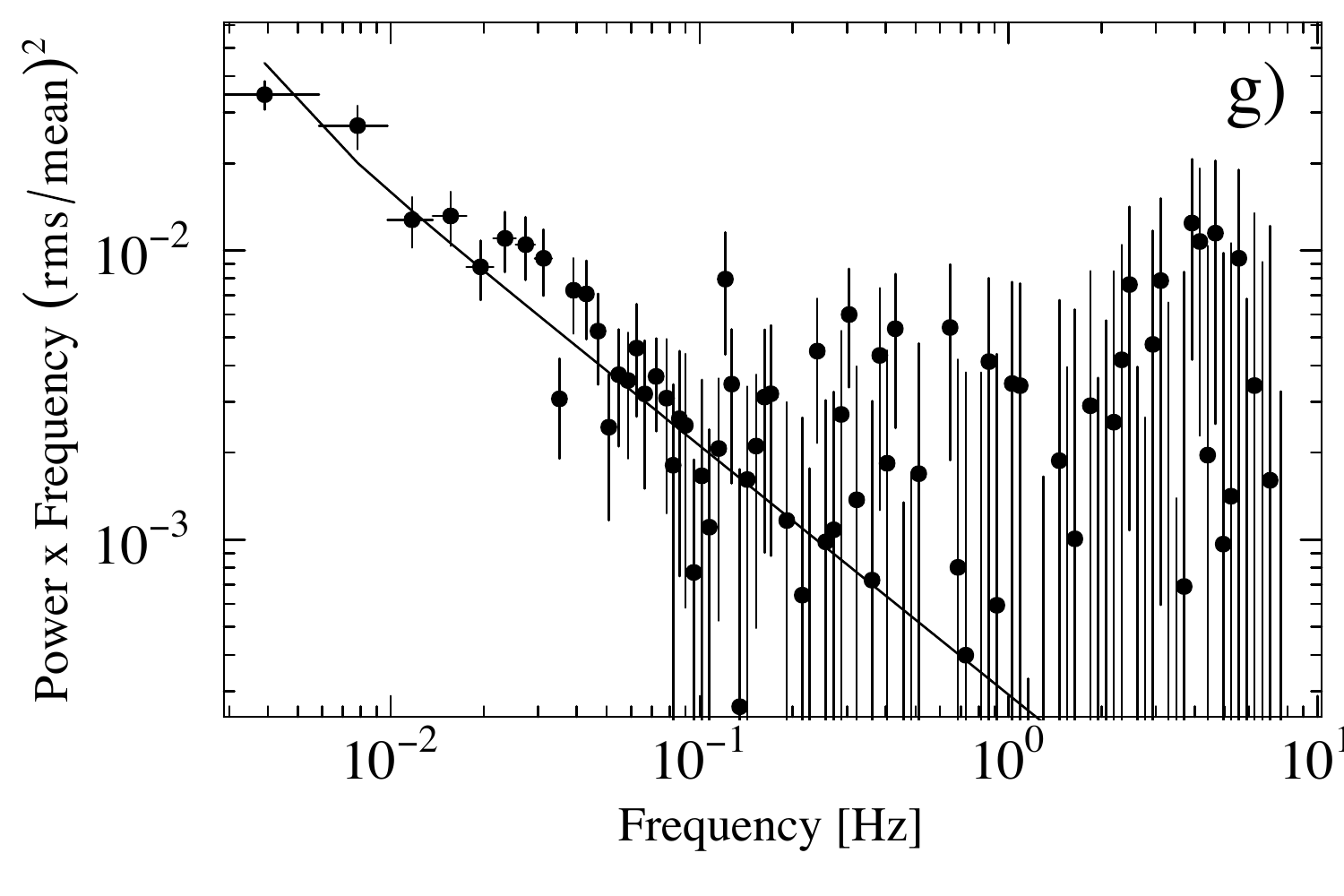}
 \caption{Pure power-law PSD from the obscured epoch labeled in Fig. \ref{asm} (\textit{bottom panel}). The Poisson noise is subtracted.}
 \label{psd_g}
\end{figure} 

The change to the obscured state occurs likely at MJD 58608 or soon after. When the source transits to the obscured phase, there is a rapid drop in $\Gamma$, log $\xi$, and flux, and a change from the highly ionized iron absorption lines to emission lines. In the obscured phase, the PSD is consistent with a pure power-law noise (Fig. \ref{psd_g}) indicating that all the intrinsic timing information is lost leaving only the scattered component with $\Gamma\sim2.0$.    

In the obscured phase, the spectral model used for fitting the data of the previous \nicer\/ pointings did not provide good fits. Instead, we used the modeling results of \citet{koljonen20}, and added a neutral lower ionization reflection component to the model (\textsc{xillver}) with all parameters tied to the \textsc{relxill} component except normalization. Thus, the model can be now described as: \textsc{vphabs} $\times$ (\textsc{relxill} + \textsc{xillver} + \textsc{lines}) + \textsc{powerlaw}. We further set the ionization parameter to 1, fixed the reflection factor to $-$2 corresponding to an obscured reflection scenario, and allowed the inclination to vary freely ( likely corresponding to some mean scattering angle of the obscuring cloud). The model parameter evolution is shown in Fig. \ref{specfit}, and the corresponding parameter values are shown in Table A2.    

\begin{figure}
 \centering
 \includegraphics[width=\linewidth]{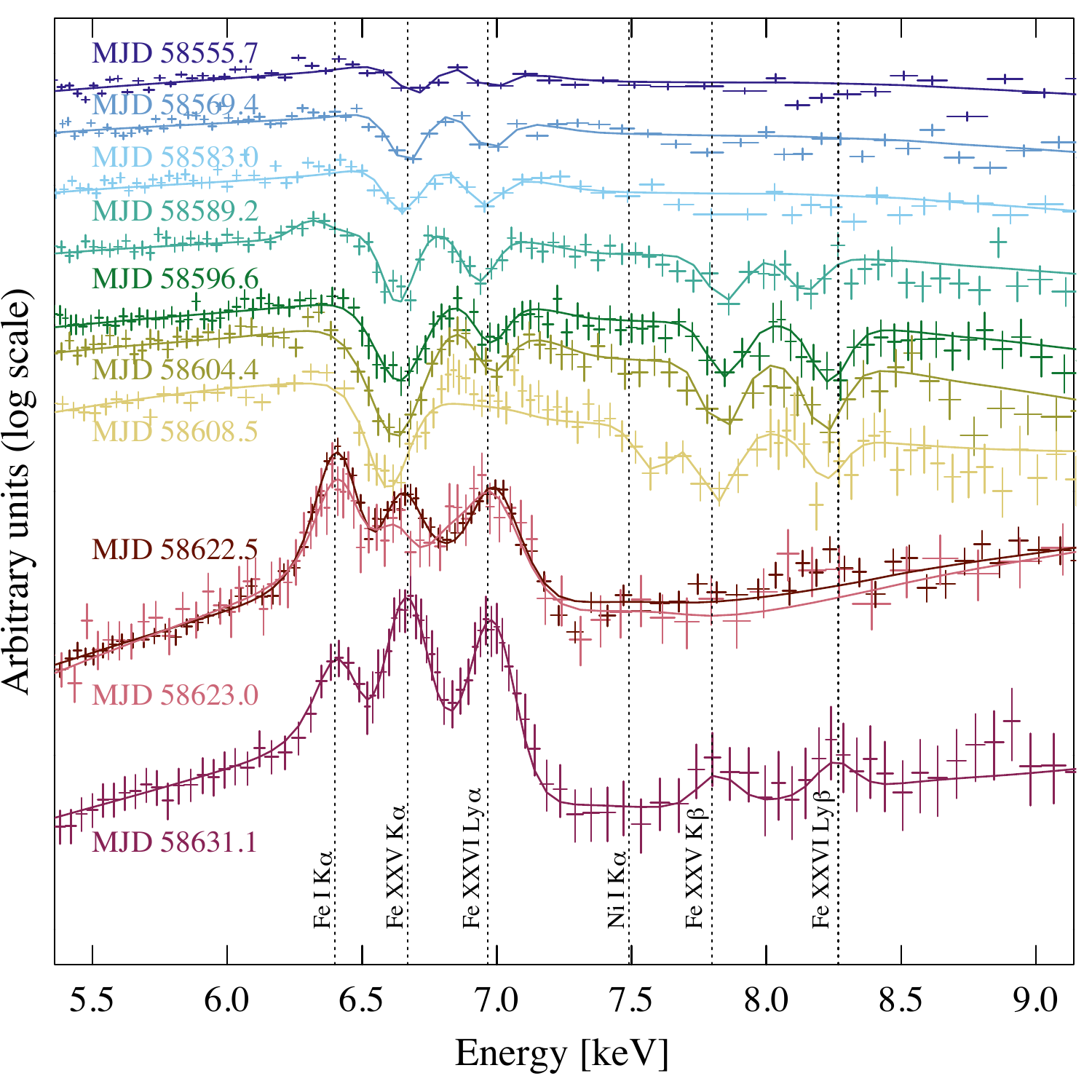}
 \caption{Evolution of the iron lines from absorption to emission when \object{GRS 1915+105} changed from the rebrightening phase (top seven spectra) to the obscured phase (bottom three spectra). Two spectra from the obscured phase (MJD 58622 and MJD 58623) are overlaid to highlight the possible absorption feature of Fe XXV K$\alpha$ line. The best-fit models are shown as solid lines.}
 \label{nicerabs}
\end{figure}

In the obscured phase, the absorption lines turn to emission together with a strong neutral iron emission line (modeled with \textsc{xillver}, but also fitted separately with a gaussian line to be able to compare values to other lines shown in Table \ref{ironlines}). The equivalent widths of the emission lines vary from $\sim$50 eV to $\sim$150 eV, most likely indicating changes in the obscuring matter. Exceptionally, in one pointing at MJD 58623, the Fe XXV K$\alpha$ line seems to be in both absorption and emission (see Fig. \ref{nicerabs}). In the previous observation, taken half a day earlier, the absorption is not visible. While the absorption of the Fe XXV K$\alpha$ line occurs preferably in the resonant line (6.70 keV), the emission is characterized by the forbidden line (6.64 keV), especially for high column densities \citep{bianchi05}. Therefore, the observation of the absorption and emission line indicates at least two scattering components during this time. Extrapolating the linear decay to the obscured state (see Fig. \ref{specfit}, bottom panel) shows that the observed flux density during the obscured state is a factor of 1.2-8.8 lower than the linear decay profile. The observations presenting the lowest fluxes during the obscured phase show a similar trend with the linear decay, but with 0.045$\times$10$^{-8}$ erg s$^{-1}$ cm$^{-2}$ of flux removed, meaning that at most approximately 80\% of the intrinsic flux density is absorbed and/or scattered in the 1--10 keV band.

\begin{table*}
\centering
\caption{Neutral and ionized absorption and emission iron line modeling results of the \nicer\/ data during the descent to the obscured state. The corresponding X-ray spectra and models are plotted in Fig. \ref{nicerabs}. The line widths of all lines except in a few cases for Fe XXV K$\alpha$ line are below detector resolution ($\sim$0.01 keV).}
\label{ironlines}
\resizebox{\linewidth}{!}{%
\begin{tabular}{lcccccccccc}
\hline
MJD: & 58555.7 & 58569.4 & 58583.0 & 58589.2 & 58596.6 & 58604.4 & 58608.5 & 58622.5 & 58623.0 & 58631.1 \\
\hline
\textbf{Fe I K$\alpha$} \\
\hline
Energy (keV) & -- & -- & -- & 6.31$^{+0.03}_{-0.04}$ & -- & -- & -- & 6.400$\pm$0.005 & 6.38$\pm$0.01 & 6.411$\pm$0.007 \\
EW (eV) & -- & -- & -- & 14$^{+5}_{-6}$ & -- & -- & -- & 138$\pm$8 & 48$\pm$6 & 90$\pm$7 \\
\hline
\textbf{Fe XXV K$\alpha$} \\
\hline
Energy (keV) & 6.70$\pm$0.03 & 6.66$\pm$0.01 & 6.65$\pm$0.20 & 6.63$\pm$0.01 & 6.63$\pm$0.01 & 6.619$\pm$0.008 & 6.59$\pm$0.01 & 6.65$\pm$0.01 & 6.64/6.70 & 6.67$\pm$0.01 \\
EW (eV) & $-$23$\pm$8 & $-$37$\pm$6 & $-$33$\pm$6 & $-$50$^{+5}_{-6}$ & $-$87$^{+9}_{-14}$ & $-$133$^{+10}_{-9}$ & $-$87$^{+11}_{-20}$ & 46$\pm$7 & 25$\pm$15/$-$29$\pm$14 & 154$\pm$9 \\
$\sigma$ (keV) & 0.01 & 0.01 & 0.01 & 0.01 & 0.07$^{+0.01}_{-0.02}$ & 0.10$\pm$0.01 & 0.06$^{+0.02}_{-0.01}$ & 0.01 & 0.01/0.01 & 0.01 \\
N (10$^{17}$ cm$^{-2}$) & 2.6 & 4.3 & 3.8 & 5.8 & 10 & 16 & -- & -- & -- & -- \\
\hline
\textbf{Fe XXVI Ly$\alpha$} \\
\hline
Energy (keV) & 6.97$\pm$0.04 & 6.97$\pm$0.02 & 6.95$\pm$0.02 & 6.93$\pm$0.02 & 6.97$\pm$0.02 & 6.98$\pm$0.02 & -- & 7.00$^{+0.02}_{-0.01}$ & 6.98$^{+0.05}_{-0.04}$ & 6.98$\pm$0.01 \\
EW (eV) & $-$17$\pm$8 & $-$26$\pm$6 & $-$27$\pm$6 & $-$36$\pm$6 & $-$32$\pm$6 & $-$33$\pm$6 & -- & 119$^{+12}_{-14}$ & 57$\pm$16 & 150$\pm$12 \\
N (10$^{17}$ cm$^{-2}$) & 3.7 & 5.7 & 5.9 & 8.0 & 6.9 & 7.3 & -- & -- & -- & -- \\
\hline
\textbf{Fe XXV K$\beta$} \\
\hline
Energy (keV) & -- & -- & -- & 7.84$\pm$0.03 & 7.84$\pm$0.02 & 7.83$\pm$0.02 & 7.81$\pm$0.02 & -- & -- & 7.82$^{+0.02}_{-0.04}$ \\
EW (eV) & -- & -- & -- & $-$37$\pm$9 & $-$43$\pm$8 & $-$54$\pm$11 & $-$55$\pm$11 & -- & -- & 33$\pm$12 \\
\hline
\textbf{Fe XXVI Ly$\beta$} \\
\hline
Energy (keV) & -- & -- & -- & 8.14$\pm$0.04 & 8.23$\pm$0.03 & 8.21$\pm$0.02 & 8.21$\pm$0.04 & -- & -- & 8.27$^{+0.03}_{-0.06}$ \\
EW (eV) & -- & -- & -- & $-$28$\pm$10 & $-$43$\pm$9 & $-$52$^{+10}_{-11}$ & $-$31$\pm$14 & -- & -- & 31$\pm$14 \\
\hline
\end{tabular}}
\end{table*}

\subsection{ALMA results}

\begin{figure}
 \centering
 \includegraphics[width=\linewidth]{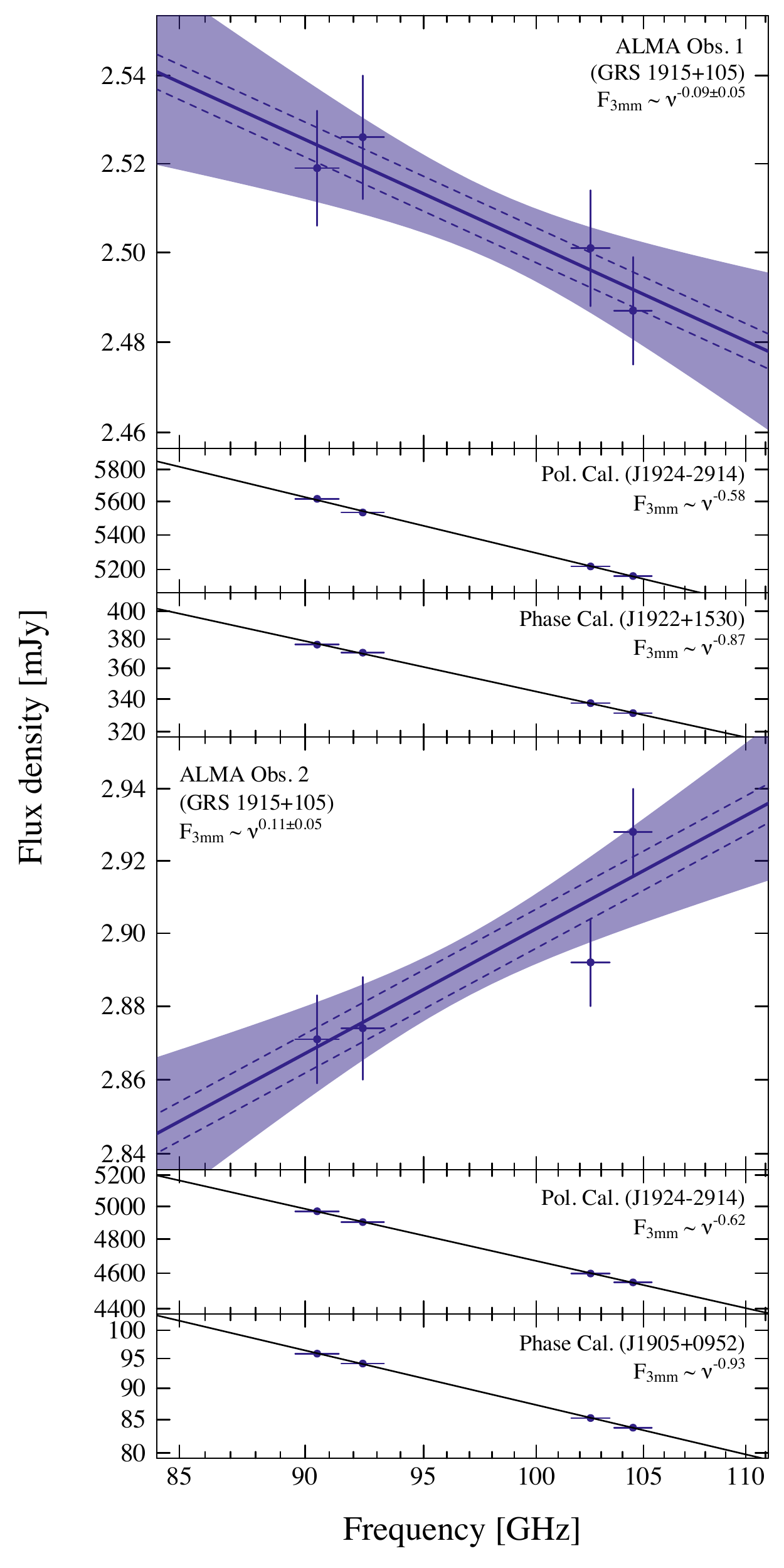}
 \caption{ALMA band 3 spectra of \object{GRS 1915+105} and calibrator sources from the two epochs.}
 \label{spectra}
\end{figure} 

The two ALMA epochs coincided with the linear decay phase and the rebrightening phase (Fig. \ref{asm}). The corresponding mean Stokes I flux densities of these epochs were 2.5 mJy and 2.9 mJy. A typical radio-quiet level for \object{GRS 1915+105} is $<$20 mJy \citep{muno01,kleinwolt02}. The flux density rise in the rebrightening phase is consistent with the rise in the X-ray flux indicating a connection between the two. As the spectral windows in the ALMA band 3 are separated into two sidebands with a 10 GHz break in between, 90--93 GHz and 102--105 GHz, we were able to estimate the in-band source spectra for the two epochs shown in Fig. \ref{spectra}. In both cases the spectra is flat with spectral indices $\alpha_{1} = -0.09\pm0.05$ and $\alpha_{2} = 0.11\pm0.05$ ($S_{\nu}\propto\nu^{\alpha}$), which are typical values for a self-absorbed synchrotron jet.  

\begin{table*}
\centering
\caption{Polarized flux densities of ALMA observations. All Stokes fluxes are shown for both epochs and spectral windows from uv-plane fitting, but the position angle and polarization fraction are only calculated for the full band. We also show the Stokes I and U from full band imaging analysis for comparison.}
\label{polares}
\resizebox{\linewidth}{!}{%
\begin{tabular}{ccccccc|cc}
& & \multicolumn{5}{c}{UVMultiFit} &\multicolumn{2}{c}{Imaging} \\
\hline
Epoch & Spw & I  & Q & U & PA & p & I & U  \\
& & (mJy) & (mJy) & (mJy) & (deg) & (\%) & (mJy) & (mJy) \\
\hline
\hspace{-0.8cm}  Ep. 1 & All & 2.499$\pm$0.007 & 0.001$\pm$0.007 & -0.031$\pm$0.006 & -44$\pm$6 & 1.3$\pm$0.2 & 2.573$\pm$0.026 & -0.040$\pm$0.014 \\
& 90.5 GHz & 2.519$\pm$0.013 & 0.004$\pm$0.012 & -0.035$\pm$0.013 \\
& 92.5 GHz & 2.526$\pm$0.014 & -0.010$\pm$0.015 & -0.055$\pm$0.015 \\
& 102.5 GHz & 2.501$\pm$0.013 & 0.005$\pm$0.012 & -0.021$\pm$0.012 \\
& 104.5 GHz & 2.487$\pm$0.012 & 0.013$\pm$0.012 & -0.059$\pm$0.012 \\
\hline
\hspace{-0.8cm}  Ep. 2 & All & 2.895$\pm$0.007 & 0.019$\pm$0.007 & -0.043$\pm$0.007 & -33$\pm$4 & 1.6$\pm$0.2 & 2.883$\pm$0.029 & -0.043$\pm$0.003 \\
& 90.5 GHz & 2.871$\pm$0.012 & 0.008$\pm$0.012 & -0.062$\pm$0.012 \\
& 92.5 GHz & 2.874$\pm$0.014 & 0.010$\pm$0.014 & -0.028$\pm$0.014 \\
& 102.5 GHz & 2.892$\pm$0.012 & 0.002$\pm$0.012 & -0.042$\pm$0.012 \\
& 104.5 GHz & 2.928$\pm$0.012 & 0.029$\pm$0.012 & -0.051$\pm$0.012 \\
\hline
\end{tabular}}
\end{table*} 

The source was weakly polarized during both epochs with polarization fractions of 1.3-1.6 \%. We could only detect significant Stokes U polarization indicating linear polarization with a position angle close to $-$45 degrees. The values for the Stokes fluxes from uv-plane fitting and imaging analysis are presented in Table \ref{polares}. We note that $\sim$1\% linear polarization is a typical value for self-absorbed synchrotron jet polarization in hard-state XRBs. The optically thick, steady jets in XRBs show low levels of linear polarization, from undetectable to a few percent \citep[e.g.,][]{corbel00,russell15}. This also agrees with earlier results from the steady jet phase with the polarization factor of the stationary core of 1-2\% \citep{fender02}.  

The jet position angle in \object{GRS 1915+105} was previously determined by tracking the jet components: -36.7$^{\circ}\pm$3.4$^{\circ}$ (northwest direction) and 146.5$^{\circ}\pm$2.8$^{\circ}$ (southeast direction; \citealt{millerjones07}), 130$^{\circ}\pm$1$^{\circ}$ \citep{reid14}, 133$^{\circ}$--157$^{\circ}$ \citep{dhawan00}, 142$^{\circ}\pm$2$^{\circ}$ \citep{fender99}, and 151$^{\circ}\pm$3$^{\circ}$ \citep{rodriguez99}. As the position angle is degenerate with $\pm$180$^{\circ}$, our values also correspond to 136$^{\circ}$--147$^{\circ}$, which agrees well with the values of the jet position angle at similar angular scales. Thus, the polarization position angles of ALMA observations are consistent with being parallel to the jet. Due to absorption effects, the intrinsic electric vector position angle of the steady jet is expected to align parallel to the magnetic field, which is expected to be parallel to the jet axis. If this is the case, this would mean that the effect of Faraday rotation would be quite small. All the above results point to the fact that \object{GRS 1915+105} exhibits a canonical compact steady optically thick XRB jet during both ALMA epochs. 

\subsubsection{Intra-epoch variability}

\begin{figure}
 \centering
 \includegraphics[width=\linewidth]{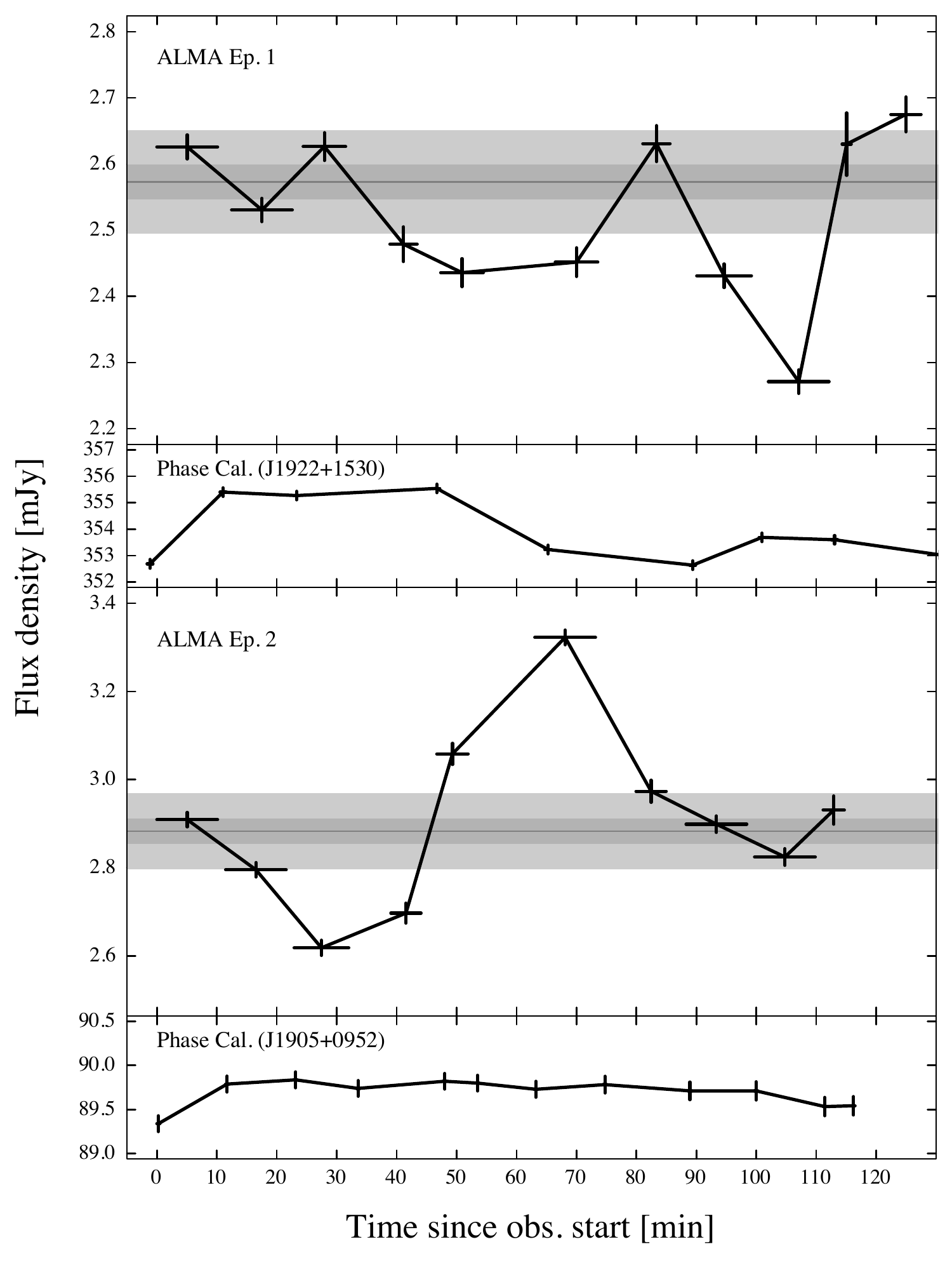}
 \caption{Band 3 light curve from the two ALMA epochs in time bins of one scan length ($\sim$10 min). The top two panels show the data from Epoch 1 and the bottom two panels from Epoch 2 for \object{GRS 1915+105} and the phase calibrators J1922$+$1530 and J1905$+$0952. The horizontal lines and the gray bands show the mean value of the flux density throughout the epochs as tabulated in Table \ref{polares}.} 
 \label{lc}
\end{figure} 

As the ALMA data consists of several scans per epoch, in addition to time-averaged analysis, we also studied the intra-epoch variability of the mm emission from \object{GRS 1915+105}. In this analysis, we considered only Stokes I flux densities because of the weakness of the polarized flux densities. The full-band Stokes I flux densities for individual scans varied between 2.3 and 2.7 mJy in Epoch 1 and 2.6 and 3.3 mJy in Epoch 2 (Fig. \ref{lc}). In the same figure, we also plot the variability of the phase calibrators. Although the variations of the phase calibrator and the target are not correlated in either epoch, the phase calibrator used during the first epoch (J1922+1530) also shows variability of similar amplitude as the target. As discussed in Sect.~\ref{almaobs}, the combined effect of a more extended array configuration and worse atmospheric conditions during the first epoch make the estimates of short-term variability less reliable. However, during the second epoch, the phase calibrator is very stable, and we conclude that the variability seen on timescales of minutes is intrinsic to the source, demonstrating usefulness  of ALMA in studying fast variability in XRBs.

\subsubsection{Millimeter/X-ray correlation}

\begin{figure}
 \centering
 \includegraphics[width=\linewidth]{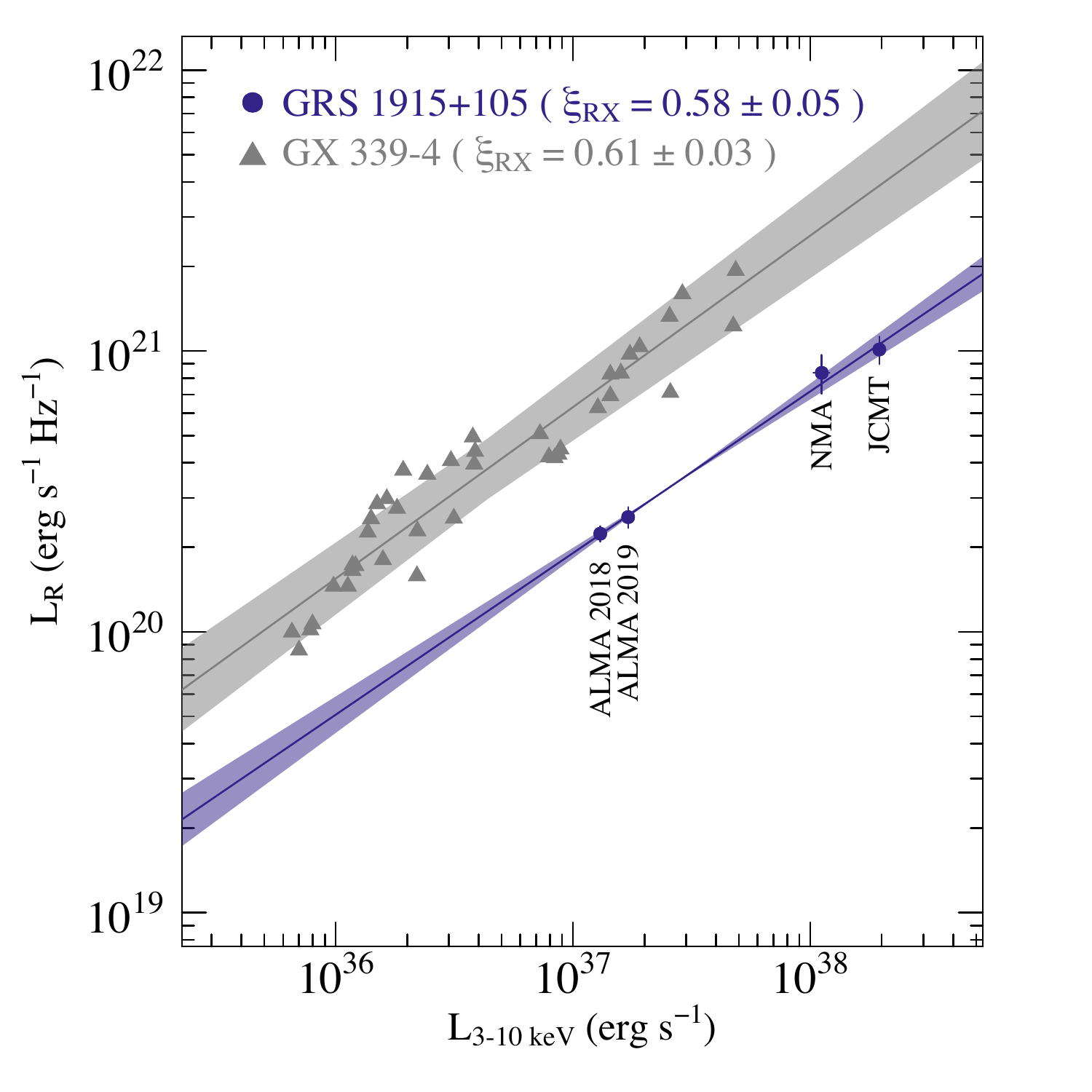}
 \caption{\object{GRS 1915+105} mm/X-ray correlation (blue points). The high-luminosity data are from JCMT \citep{ogley00} and NMA \citep{ueda02} together with quasi-simultaneous RXTE observations. In comparison, GX 339-4 hard-state radio/X-ray correlation is shown (gray triangles; same X-ray band, but radio luminosity is from 8.4 GHz; from \citealt{koljonen19}). Both have similar correlation slopes (assuming flat spectrum), reinforcing the hard-state nature of \object{GRS 1915+105} during ALMA observations.}
 \label{rxplot}
\end{figure}   

Radio/X-ray correlation is one of the most important pieces of observational evidence in connecting the mass accretion rate onto the compact object during an outburst event to the mass-loading of the jet that seems to be at work in both XRBs \citep[e.g.,][]{hannikainen98,corbel03,gallo03} and AGNs \citep[e.g.,][]{merloni03,falcke04}. During low-luminosity hard X-ray states in XRBs, the logarithmically scaled X-ray and radio luminosities present a tight relation of $L_{\mathrm{radio}} \propto L_{\mathrm{X}}^{\sim0.7}$. This relation can be explained by assuming that the X-ray emission arises from Compton scattering in advection-dominated accretion flow (ADAF) and the radio emission from the optically thick synchrotron emission in the jet \citep{heinz03}.        

To investigate the radio/X-ray correlation of \object{GRS 1915+105}, we searched for millimeter and X-ray observations during radio-quiet/plateau states to compare them with our ALMA data. We used the 350 GHz James Clerk Maxwell Telescope (JCMT) observations from \citet{ogley00} that were taken during a radio-quiet state and 94 GHz Nobeyama Millimeter Array (NMA) observations from \citet{ueda02} that were taken during a radio plateau state, both of which should present a compact steady jet. The corresponding 3-10 keV X-ray fluxes were estimated using quasi-simultaneous archival RXTE data taken from HEASARC and reduced according to standard procedures. The resulting mm/X-ray correlation can be seen in Fig. \ref{rxplot} with a correlation coefficient of 0.58$\pm$0.05. For comparison, we plotted the GX 339-4 hard-state radio/X-ray correlation (gray triangles; same X-ray band, but radio luminosity is from 8.4 GHz; from \citealt{koljonen19}). Both have similar correlation coefficients (assuming flat spectrum), reinforcing the hard-state nature of \object{GRS 1915+105} during ALMA observations and the decay phase.

\section{Discussion}

To briefly summarize the above results, we find that \object{GRS 1915+105} exhibits typical XRB hard-state properties during the exponential and linear decay phases. These include decreasing X-ray luminosity down to $\sim$1\% of the Eddington luminosity, PSD with a band-limited noise profile and a type-C QPO with a harmonic, an absorbed power-law spectrum that can be fitted with a Comptonization model, and a weakly polarized, compact, optically thick jet. The mm/X-ray correlation shows similar coefficient to that of canonical low-luminosity XRBs in the radiatively inefficient track, further suggesting that the source is in the canonical hard state. The X-ray light curve profile showing first an exponential decay followed by a linear decay is a hallmark of a viscous and irradiation-controlled decay observed during the end stages of transient XRB outbursts \citep[e.g.,][]{tetarenko18}. However, the following rebrightening phase displayed similar X-ray and radio properties to the preceding exponential and linear decay phases, with elevated X-ray and radio luminosity marking a departure from the linear decay trend. In addition, the PSDs show an additional red noise power-law component increasing in rms, and the X-ray spectra show high-ionization absorption lines and modeling results indicate increasing absorption. When the source transits to the obscured state, the X-ray properties change markedly with an order of magnitude drop in the X-ray flux, the PSDs show a pure red noise profile, and the X-ray spectra becomes much harder with prominent high-ionization emission lines.

\subsection{Comparison with the previous low X-ray flux states}

During the outburst, we divided the low-X-ray-flux states of \object{GRS 1915+105} into two different states depending on the strength of the radio flux density: a radio-quiet hard ($\chi$) state with a radio flux density of a few mJy, and a plateau state with the radio flux density presenting higher, typically 50-100 mJy flux densities \citep{foster96,pooley97,fender99}. The former is close to a `normal' XRB hard state, while the latter is probably a high-luminosity excursion from a very high state. For both states, the jet properties are similar (apart from the flux density) and correspond to a steady, flat-spectrum, optically thick compact jet. Selecting observations at the plateau state which  presents a compact steady jet, \citet{rushton10b} were able to find a radio/X-ray relation of $L_{\mathrm{radio}} \propto L_{\mathrm{X}}^{\sim1.7\pm0.3}$ or $L_{\mathrm{radio}} \propto L_{\mathrm{X}}^{\sim1.1\pm0.1}$ if  only taking the coronal emission 
into account when estimating the X-ray luminosity \citep{peris16}. 

Many XRBs show a steeper radio/X-ray correlation during high-luminosity phases of the hard state \citep[e.g.,][]{coriat11}. It is unclear whether the change depends on the source being in outburst rise or decline \citep{koljonen19,islam18} or there exists multiple correlations depending on the physical qualities of the systems \citep{gallo12}. The different physical mechanisms for multiple correlations have been suggested to arise from differences in radio emission properties \citep{casella09,espinasse18}, differences in X-ray emission properties \citep{meyerhofmeister14,coriat11,xie16}, or inclination effects on X-ray emission properties \citep{heil15,munosdarias13,motta18,petrucci01,niedzwiecki05} and radio emission properties \citep{soleri11,motta18}. Whatever the case, the change of the correlation coefficient of \object{GRS 1915+105} to $L_{\mathrm{radio}} \propto L_{\mathrm{X}}^{\sim0.6}$ indicates a change in the evolution of the outburst to a low-luminosity hard state.

Both plateau and $\chi$ states show similar X-ray timing properties to those observed during the exponential and linear decay with a PSD presenting a flat-top noise profile and a type-C QPO. However, the radio-quiet state shows QPOs with central frequencies greater than $\sim$2 Hz while the plateau state shows QPOs with central frequencies lower than $\sim$2 Hz \citep{pahari13}. Interestingly, similar behavior is seen in the transition from exponential and linear decay phases that display a transitional QPO frequency of  2 Hz. In addition, it has been found that the average QPO phase lag is positive above and negative below 2 Hz \citep{reig00,lin00,qu10,pahari13,zhang20}, and a similar trend can be observed between the exponential and linear outburst decay phases. This effect has been attributed to a pivoting lag-energy spectrum with the lag increasing with energy for progressively lower QPO frequencies and decreasing for progressively higher QPO frequencies \citep{pahari13,zhang20}. 

Several explanations for the phase lag behavior have been discussed in the literature. \citet{vandeneijnden17} relate the phase lags to the inclination of the system. Soft lags are observed from high-inclination systems, and hard lags are observed from low-inclination systems for QPO frequencies exceeding 2 Hz. In particular, for \object{GRS 1915+105}, the phase lag behavior is linked to an evolving two-temperature corona \citep{nobili00} or two precessing regions of the accretion flow \citep{vandeneijnden16}. While differing in physical picture, the underlying assumption in both scenarios is the same, presenting two (or more) regions located at different radii and having a different spectral response. The difference in the location produces the change in the QPO frequency, while the different spectral response produces the change in the phase lag. Effectively, this means that for high QPO frequencies the spectral response of the inner part of the accretion flow is harder than that of the outer part, while for low QPO frequencies the outer part presents a harder response. The zero time lag with $\sim$2 Hz QPO frequency would correspond to a similar spectral response from both regions.

During the transition to the linear decay phase, the lag at the QPO frequency decreases, indicating that the dominating emission region has moved outwards. Relating this behavior to the plateau/radio-quiet state of the outburst phase means that the plateau is a hard X-ray state with a compact hot inner flow while the radio-quiet state presents more expanded hot inner flow. However, the behavior of the radio emission is different, with the radio flux density staying at a low level of 1-3 mJy throughout the linear decay phase \citep{motta21} while the plateau state presents much higher radio flux densities. However, this can be expected, as the mass accretion rate during the outburst phase is at least an order of magnitude higher (see Fig. \ref{hid}).

\subsection{Viscous outburst decay?}

Assuming the observed light-curve profile obeys the standard XRB outburst decay profile, we fit the \nicer\/ light curve with a model used in \citet{powell07}, \citet{heinke15}, and \citet{tetarenko18} corresponding to an exponential decay on a viscous timescale and a linear decay on an irradiation-controlled timescale. The full model can be written as: 

\begin{equation}
  f_{X} =
  \begin{cases}
    (f_{t}-f_{2}) \, \mathrm{exp} \, (-(t-t_{break})/\tau_{e})+f_{2} & \text{if $t \leq t_{b}$} \\
    f_{t}(1-(t-t_{b})/\tau_{l}) & \text{if $t > t_{b}$}
  \end{cases}
,\end{equation}

where $\tau_{e}$ is the viscous (exponential) timescale, $\tau_{l}$ is the irradiation-controlled (linear) timescale, $t_{b}$ is the transition time, $f_{t}$ is the flux density at the transition, and $f_{2}$ is the asymptotic flux density of the exponential decay. The fit resulted in the following parameters: $\tau_{e}=58\pm4$ days, $\tau_{l}=461\pm6$ days, $t_{b}=$ 58328$\pm$1 days (MJD), $f_{t}=0.195\pm0.001 \times$ 10$^{-8}$ erg s$^{-1}$ cm$^{-2}$, and $f_{2}=0.07\pm0.02 \times$ 10$^{-8}$ erg s$^{-1}$ cm$^{-2}$.

Taken at face value, the timescale of the exponential decay (58 days) appears to be relatively low. The approximate viscous time of the whole disk is at least $t_\mathrm{visc} \approx \alpha^{-1} (H/R)^{-2} t_\mathrm{dyn} \sim 500$ days, where we take the viscosity parameter as $\alpha=0.2$, the scale height to radius ratio as $H/R=0.1,$ and $t_\mathrm{dyn}\sim1$ day as the reciprocal of the Keplerian frequency at the circularization radius ($\sim$2$\times$10$^{12}$ cm). However, the average viscosity parameter of the whole disk is likely much lower as only the irradiated part of the disk is in the hot state, which would make the viscous timescale even longer. Possible mechanisms to decrease the viscous time include increasing the value of the average viscosity parameter and/or increasing the scale height to a radius ratio from the canonical values. In addition, the outflowing wind reduces the viscous timescale by a factor of $(1-e_{w}),$ which for the strong disk wind of \object{GRS 1915+105} can be a sizable effect.  

Typically, observation of the exponential decay requires that the disk has been completely ionized and $R_{\mathrm{irr}}>R_{\mathrm{out}}$. As \object{GRS 1915+105} harbors such a large disk, this is likely not the case. Instead, it appears that the outburst decay behaves as if \object{GRS 1915+105} has only a disk size of $R_{\mathrm{out}}=R_{\mathrm{irr}}\sim5\times10^{11}$ cm. Alternatively, the exponential decay may arise from another process that reduces the mass accretion rate to the black hole. \citet{cannizzo00} discusses a case where the evaporation of the thin accretion disk into a hot inner flow close to the compact object is dominant over the viscous evolution. Strong evaporation leads to an e-folding decay rate associated with the loss of material from the inner accretion disk that can be ten times faster than the viscous evolution. Indeed, the critical luminosity for the disk evaporation scheme to hold is considered to be approximately 5\%\ of the Eddington luminosity \citep{meyerhofmeister04}, which matches with the luminosities observed at the end of the exponential decay.

\subsection{Geometrical implications of the outburst decay} 

Based on the canonical scenario, the X-ray spectra showing a slowly hardening and diminishing Comptonization component corresponds to the reducing radiative cooling of the optically thin and hot inner accretion flow as the optically thick accretion disk recedes or evaporates to larger radii. If the power-law component in the PSD during the exponential decay is flicker noise (with $\Gamma\sim$1) from the accretion disk \citep{lyubarskii97}, this supports the receding or evaporating disk. However, a disk component is not statistically needed to fit the X-ray spectra, but as the absorption towards \object{GRS 1915+105} is high, the modeling of the curved soft X-ray part of the spectra can be degenerate and a low-temperature disk could still produce some of the soft X-rays. Similarly, the rms of the zero-centered Lorentzian increases from 10\% to 20\%, which implies that the dilution from the disk is not impacting the rms in the linear decay phase. In the linear decay phase, the spectral evolution stops and settles into a stable state with slowly decreasing flux density and QPO frequency. If the type-C QPO frequency is tied to the size of the precessing hot inner flow \citep{ingram09}, this means that the disk is still receding, but does not present enough cooling of the inner flow for the spectral evolution to remain constant. 

In the rebrightening phase, the evolution of both the spectral and timing properties have departed from the linear decay. The X-ray power-law photon index and ionization parameter from the continuum modeling, in addition to the X-ray and radio flux densities, present elevated values. Thus, it seems that the mass accretion rate has increased slightly to the compact object. At the same time, the X-ray spectra show high-ionization absorption lines and the power-law component steepens and becomes prominent in the PSD during the latter part of the rebrightening phase, indicating an increased amount of absorbing and scattering material along the line of sight. The redshifted Fe XXV K$\alpha$ could arise from inflowing dense clumps in the disk atmosphere ans/or corona \citep{kubota18}. However, this scenario does not explain why Fe XXVI Ly$\alpha$ line is not redshifted. Alternative to a redshifted line, several ionization zones that produce several absorbing charge states can broaden the absorption line without invoking a fast wind \citep{miller20}. This scenario would indicate a relatively dense medium of N$_{\mathrm{H}}=$3--6$\times$10$^{23}$ cm$^{-2}$ in the line of sight. 

Based on \chandra\/ data, \citet{miller20} argued that the obscuration is caused by a ``failed wind'' that arises close to the black hole and is unable to escape from the system. Their first\chandra\/ observation coincided with the latter part of the rebrightening phase (MJD 58603) where there was evidence of obscuration through photoionized absorption. Alternatively, the emerging obscuration could arise in a radially stratified, puffed-up outer disk as discussed in \citet{neilsen20}. Based on photoionization modeling of \nicer\/ spectra during an X-ray flare in the obscured state, these latter authors argued that the X-ray absorption takes place further out in a vertically extended outer disk (R$\sim$few$\times$10$^{11}$ cm). The puffed-up outer disk could either arise from changes in the inner accretion flow increasing the temperature and scale-height of the outer disk by irradiation, or a structural change of the outer disk connected to the ending of the outburst and a switch to the quiescent state. 

Based on these scenarios and our modeling of the \nicer\/ data, the failed wind or puffed-up outer disk would gradually form during the rebrightening phase. This can be seen both in the continuum modeling, in the absorption line evolution, and in the X-ray PSD evolution. In the continuum, local absorption from lighter elements, as well as the increase of the reflection factor, indicates increasing scattering in the surrounding matter with time. At the same time, the equivalent widths of the highly ionized iron absorption lines increase, and the ratio of Fe XXVI/Fe XXV decreases, indicating the presence of increased cooler material in the line of sight. Also, the steepening power-law component in the PSD as well as the dilution of the rms of the zero-centered Lorentzian is consistent with a scattering medium slowly entering the line of sight. The additional luminosity seen in the rebrightening phase with respect to the linear decay could arise from reaccreting the failed wind or some interaction with the wind, such as for example backscattering of emission from the far side of the scattering cloud, or a structural change of the inner accretion flow resulting in an increase in the mass accretion rate. Although the X-ray observations may indicate an emerging wind in the rebrightening phase, which in principle could work as a depolarizing medium for the mm emission from the jet, the polarization fraction of the mm emission increases in the rebrightening phase. This could either mean that the depolarization of the putative wind does not present a measurable effect on the polarization or that the mm emission line of sight does not coincide with the disk wind.
       
The change to the obscured state occurs likely at MJD 58608 or soon after. The spectra during the \nicer\/ pointing still show highly ionized absorption lines but present a clear drop in the photon power law index, ionization parameter, and flux density of the continuum departing from the levels of the rebrightening phase. The rest of the \nicer\/ pointings are well into the obscured phase. Extrapolating the flux decay from the linear decay phase shows that $\sim$80\% of the observed flux is absorbed or scattered. 

\subsection{The fate of the \object{GRS 1915+105} outburst}

The fate of the \object{GRS 1915+105} outburst is currently unclear. Nevertheless, a return to a ``regular'' hard state, not been seen before during the outburst, can be established based on the present work, as detailed above. If the outburst of GRS 1915+105 is taken to be similar to other XRBs, a return to a hard state heralds the impending end of the outburst and an eventual return to quiescence. On the other hand, recent work on the obscured state by \citet{balakrishnan21}, \citet{motta21}, and \citet{neilsen20} shows that \object{GRS 1915+105} has exhibited both strong X-ray and radio flaring, indicating a significant mass accretion rate to the compact object. However, due to the heavily modified X-ray spectra by the obscuring material, the intrinsic X-ray luminosity is difficult to estimate accurately with values ranging from 1\% \citep{koljonen20,miller20} to 10\% \citep{balakrishnan21,neilsen20} and possibly reaching 100\% of the  Eddington ratio occasionally \citep{balakrishnan21}. In addition, it is not altogether clear whether the radiative efficiency of the jet remains constant. If there is matter expelled in the direction of the jet, it can interact with the jet material producing shocks and more efficient dissipation of the jet kinetic energy. 

Possible clues as to the underlying mass accretion rate may come from the much softer state observed during September 2020. Based on \maxi\/ data \citep[][see also Fig. \ref{hid}]{motta21}, the X-ray spectrum can be fitted with a thermal model with a flux corresponding to the level observed in the linear decay phase (assuming no obscuration). Excursions to lower hardness ratios during the outburst decays of XRBs are seen, for example during the 2002/2003 outburst of GX 339$-$4 \citep{belloni05}. On the other hand, it is not clear in what way the softer state during September 2020 is affected by absorption and the source could also present larger intrinsic luminosities \citep{motta21}.

As discussed in \citet{miller20} and \citet{motta21}, \object{GRS 1915+105} may be experiencing a phase of high obscuration under which the source continues to accrete at the same rate as before. At some point in the future, when the obscuring matter is lifted, we will see the rise in the apparent luminosity, which could be what happened in 1992 at the ``onset'' of the outburst. On the other hand, given the results presented above, it is reasonable to assume that \object{GRS 1915+105} has reached a low-luminosity hard state and could be on its way to quiescence. As the accretion disk of \object{GRS 1915+105} is very large, we are seeing the return to quiescence play out in slow motion as compared to other XRBs with much smaller disks, and this process can therefore take years. 

\subsection{Implications of an $\sim$30 year outburst} \label{disc1}

In this section, we entertain the possibility that the outburst began in 1992 and that the source is heading towards quiescence in the near future, and what that would imply in terms of system parameters.

Truss \& Done (2006) give estimates on the outburst duration based on the assumed amount of matter accreted during the whole outburst. These latter authors estimate that for the disk size of $R_{\mathrm{disk}}=27\times10^{11}$ cm the outburst duration would exceed a thousand years assuming that all the matter in the disk is accreted and that the surface density throughout the disk is equal to the critical surface density required to trigger an outburst through the thermal-viscous instability. However, using a more realistic surface density profile and a mass loss due to a disk wind, Truss \& Done (2006) were able to reduce this to 76-160 years (where the range comes from assuming zero or Eddington mass loss for the disk wind). This timescale is still too long for a 30-year outburst. 

The duration can be further decreased assuming that only a part of the disk participates in the outburst. Truss \& Done (2006) give a range of $t_\mathrm{outburst}$/(1+$e_{w})=47-23$ years depending on the mass loss of the disk wind $e_{w}=\dot{M}_{w}/\dot{M}_{\mathrm{Edd}}=0-1,$ and using a radius of influence of the incident X-rays of $R_{\mathrm{irr},11}=15,$ which fits with the assumed duration of 30 years with $e_{w}\sim0.75$. The average mass loss rate of the wind has been estimated to be approximately $\dot{M}_{w} \approx 10^{19}$ g s$^{-1}$ \citep{neilsen12,miller16}. Assuming that during the outburst \object{GRS 1915+105} accreted close to the Eddington accretion rate implies a mass accretion rate of $\dot{M}_{\mathrm{Edd}}=1.7\times10^{19}$ g s$^{-1}$ for a 12 solar mass black hole,  giving $e_{w}=\dot{M}_{w}/\dot{M}_{\mathrm{Edd}}\approx0.6,$ which is compatible with the value estimated from the outburst duration.  

A more rigorous study with smooth particle hydrodynamics simulations including the effects of the thermoviscous instability, tidal torques, irradiation by central X-rays, and wind mass-loss resulted in a similar outburst length of $t_\mathrm{out}=20-40$ yr for an irradiation efficiency of $\epsilon \sim 10^{-3}$ \citep{deegan09}. \citet{truss06} give a slightly larger value for the irradiation efficiency of $\epsilon = 1.6 \times 10^{-3}$ for the above radius $R_{\mathrm{irr}}$ and given Eddington luminosity and a standard radiation efficiency of $\eta=0.1$. However, both studies concluded that the irradiation efficiency (on average) cannot be higher as it would make the outburst length much longer.

Based on the assumed outburst duration and the small value of the irradiation efficiency, it seems clear that only a part of the disk has participated in the outburst. In the following, we estimate the fraction of the disk mass accreted during the outburst of \object{GRS 1915+105}. The mass accreted during the outburst can be estimated from the time-averaged rate of mass loss from the accretion disk: 

\begin{equation} \label{eq2}
\langle \dot{M}_{\mathrm{disk}} \rangle = \langle \dot{M}_{\mathrm{Edd}} \rangle + \langle \dot{M}_{w} \rangle - \dot{M}_{2}, 
\end{equation}

where values for $\dot{M}_{\mathrm{Edd}}$ and $\dot{M}_{w}$ are given above. The mass-transfer rate from the companion $\dot{M}_{2}$ can be estimated from \citet{deegan09} and \citet{ritter99}:

\begin{equation} \label{eq3}
-\dot{M}_{2} \sim 7.3 \times 10^{-10} \, \Bigg( \frac{M_{2}}{\mathrm{M}_{\odot}} \Bigg)^{1.74} \Bigg( \frac{P_{\mathrm{orb}}}{1 \mathrm{d}} \Bigg)^{0.98} \, M_{\odot} \, \mathrm{yr}^{-1},
\end{equation}

which gives $-\dot{M}_{2} \sim 1.3\times10^{18}$ g s$^{-1}$ for system parameters of \object{GRS 1915+105}. Alternatively, we can use the modeled asymptotic flux density of the exponential decay fitted to the light curve, $f_{2}=0.7\times10^{-8}$ erg s$^{-1}$ cm$^{-2}$, which for the distance of 8.6 kpc and a standard radiative efficiency of $\eta=0.1$ corresponds to $\sim$7$\times10^{18}$ g s$^{-1}$ when taking into account a bolometric correction of 10, which is typical for a hard-state XRB \citep{koljonen19}. This is of similar order to the value estimated from Eq. \ref{eq3}. Inserting $-\dot{M}_{2}=1.3\times10^{18}$ g s$^{-1}$ to Eq. \ref{eq2} results in $\dot{M}_{\mathrm{disk}}\sim2.6\times10^{19}$ g s$^{-1}$ , which for the assumed outburst duration of 30 years is 2.3$\times10^{28}$ g mass lost. To regain this mass via the mass-transfer rate from the companion would take approximately 560 years. 

An approximate or upper limit for the total mass of the disk before the onset of the outburst can be roughly estimated from: 

\begin{equation} \label{eq4}
M_\mathrm{disk} = \int_{0}^{R_\mathrm{out}} 2 \pi R \Sigma (R) dR, 
\end{equation}

and assuming that the surface density, $\Sigma$, at all radii, $R$, is equal to the critical surface density, $\Sigma_{\mathrm{max}}$, to trigger an outburst via the thermal-viscous instability \citep{hameury98}: 

\begin{equation} \label{eq5}
\Sigma_{\mathrm{max}} = 13.4 \alpha_{c}^{-0.83} \Bigg( \frac{M_{1}}{M_{\odot}} \Bigg)^{-0.38} \Bigg( \frac{R}{10^{10} \, \mathrm{cm}} \Bigg)^{1.14} \, \, \mathrm{g} \, \mathrm{cm}^{-2},
\end{equation} 

where $\alpha_{c}$ is the cold viscosity parameter. Inserting Eq. \ref{eq5} to Eq. \ref{eq4}, and using $\alpha_{c}=0.02$ and $M_{1}/M_{\odot}=12$ gives $M_\mathrm{disk}$ = 2.7$\times10^{22} \, R_{10}^{3.14}$ g. Assuming that the disk can at least  reach the circularization radius, which for \object{GRS 1915+105} is approximately 2$\times10^{12}$ cm, gives $M_\mathrm{disk} = 4.5\times10^{29}$ g, resulting in 5\% of the total mass being accreted or lost during outburst. Similarly, the radius which corresponds to the lost disk mass is approximately $R=8\times10^{11}$ cm.

Using the irradiation law from \citet{dubus01}, $T^{4}_{\mathrm{irr}}=\epsilon L_{\mathrm{Edd}}/4\pi\sigma_{\mathrm{SB}}R_{\mathrm{irr}}^{2}$, the irradiated temperature needed to fully ionize an accretion disk layer, $T_{H}\sim$10$^{4}$ K \citep{dubus99}, the irradiation efficiency of $\epsilon=10^{-3}$, and Eddington luminosity for $L_{\mathrm{bol}}=1.5\times10^{39}$ erg s$^{-1}$ cm$^{-2}$ results in $R_{\mathrm{irr}}\sim5\times10^{11}$ cm for the irradiation radius, which is much less than the size of the disk ($\sim$2$\times10^{12}$ cm), but close to the radius where mass equates the mass lost from the disk in the outburst as determined above. Thus, assuming the outburst is nearing its end, we can conclude that the disk region that participated in the outburst is bounded by the irradiation radius, which is on the order of 5--8$\times10^{11}$ cm, and contains about 5\% of the total disk mass.   

\section{Conclusions}

We conducted two full polarization ALMA observations together with almost daily \nicer\/ pointing observations to study the jet and accretion disk properties during the outburst decay in 2018--2019 (Sect. 2). We divided the outburst decay into four distinct phases: an exponential decay phase, a linear decay phase, a rebrightening phase, and an obscured phase (Sect. 3.1). The first two phases commonly occur during a decaying XRB outburst, and we show that the X-ray spectral and timing properties of GRS 1915+105 are indeed typical for a decaying XRB outburst (Sect. 3.2). In addition, the jet emission in the mm is consistent with a compact, steady jet showing $\sim$1\% linear polarization, and the magnetic field likely aligned with the jet position angle (Sect. 3.3). Together with archival mm observations in the hard state, we formed a mm/X-ray correlation that revealed a correlation coefficient of 0.6 between the logarithmically scaled luminosities (Sect. 3.3.2). The latter two decay phases are anomalous and present evidence of increased absorption and scattering (Sects. 3.2.3 and 3.2.4) likely in the form of an accretion disk wind or a puffed-up outer disk (Sect. 4.3).

Due to the large mass reservoir of the accretion disk in \object{GRS 1915+105}, and assuming the outburst is ending in the near future, the source would have managed to accrete or eject only a small part of the matter available in the disk during its three-decade-long outburst (Sect 4.5). The relatively short outburst duration also requires both a strong disk wind and a small irradiation efficiency leading to the irradiated part of the disk being much less than the size of the disk. This is in direct discrepancy with the exponential decay profile, which is typically linked to the viscous decay of fully irradiated disks. We speculate that efficient evaporation of the inner accretion disk could be responsible for the e-folding decay profile (Sect 4.2). Of course, all of these problems could be solved by assuming that the outburst is still continuing and we have only witnessed a peculiar transit first to a canonical low-luminosity hard state and secondly to a heavily obscured but intrinsically bright accretion state (Sect 4.4).

Since the beginning of its outburst, \object{GRS 1915+105} has shown remarkable behavior in emission from the accretion disk and the jet, and the recent observations of a variety of new accretion states show no exception. The peculiarities likely arise from the large disk size and our near-edge-on viewing angle to the disk, allowing us to study the geometrical effects of the accretion flow. In addition, due to the long viscous time of the accretion disk in \object{GRS 1915+105}, we might also be witnessing events in slow motion as compared to much smaller disks in other XRBs, which enhances the importance of detailed studies of \object{GRS 1915+105} in the near future.   

\begin{acknowledgements}
K. I. I. K. was supported by the Academy of Finland project 320085.
T. H. was supported by the Academy of Finland projects 317383, 320085, and 322535. We thank the anonymous referee for useful comments. In addition, we thank the staff of the Nordic ARC Node for help in analyzing the reliability of the fast variability in our ALMA data. This paper makes use of the following ALMA data: ADS/JAO.ALMA\#2018.1.00621.S. ALMA is a partnership of ESO (representing its member states), NSF (USA) and NINS (Japan), together with NRC (Canada), MOST and ASIAA (Taiwan), and KASI (Republic of Korea), in cooperation with the Republic of Chile. The Joint ALMA Observatory is operated by ESO, AUI/NRAO and NAOJ. This research has made use of data and/or software provided by the High Energy Astrophysics Science Archive Research Center (HEASARC), which is a service of the Astrophysics Science Division at NASA/GSFC.
\end{acknowledgements}

\bibliographystyle{aa}

\bibliography{references}

\begin{appendix}

\longtab[1]{
{\small\tabcolsep=3pt  
\begin{longtable}{lllllllllllll}
\caption{\nicer\/ timing analysis parameters.} \\
\toprule
\multicolumn{3}{l}{\textbf{Exponential decay phase}} \\
\midrule
& & & Model: & (P)owerlaw & \multicolumn{2}{l}{(L)orentzian (0)} & \multicolumn{3}{l}{Lorentzian (1) = QPO Main freq.} & \multicolumn{3}{l}{Lorentzian (2) = QPO 1st harmonic} \\
\nicer\/ & Date & Exp. & & $\Gamma$ & $\sigma_{0}$ & rms$_{0}$ & $\nu_{1}$ & Q$_{1}$ & rms$_{1}$ & $\nu_{2}$ & Q$_{2}$ & rms$_{2}$ \\
obs. & MJD & (s) & & & (Hz) & (\%) & (Hz) & & (\%) & (Hz) & & (\%) \\
\midrule
A137 & 58238.609 & 5491 & PL012 & 0.8$^{+0.1}_{-0.1}$ & 6$\pm$1 & 9$\pm$1 & 4.18$\pm$0.04 & 4.1$^{+0.7}_{-0.6}$ & 6.3$\pm$0.4 & 8.8$^{+0.3}_{-0.2}$ & 4$^{+3}_{-2}$ & 3.4$\pm$0.7 \\
A138 & 58238.998 & 1935 & PL012 & 1.0$^{+0.6}_{-0.3}$ & 5$_{-1}^{+2}$ & 10$\pm$1 & 3.86$\pm$0.05 & 5$\pm$1 & 6.6$\pm$0.6 & 7.9$\pm$0.5 & 2$^{+3}_{-1}$ & 5$^{+1}_{-2}$ \\
A139 & 58243.886 & 1386 & L012 & -- & 7$\pm$1 & 13$\pm$1 & 3.56$\pm$0.04 & 7$^{+2}_{-1}$ & 6.9$\pm$0.6 & 7.2$^{+0.2}_{-0.3}$ & 7$^{+11}_{-5}$ & 4$\pm$1 \\
A140 & 58244.015 & 1440 & PL012 & 0.8$^{+0.7}_{- 0.3}$ & 6$^{+1}_{-2}$ & 12$^{+1}_{-2}$ & 3.36$\pm$0.04 & 5$^{+2}_{-1}$ & 7.3$^{+0.8}_{-0.7}$ & 6.8$\pm$0.2 & 5$^{+4}_{-2}$ & 5$\pm$1 \\
A142 & 58260.064 & 1038 & L01 & -- & 9$\pm$1 & 16.4$^{+0.7}_{-0.7}$ & 3.09$\pm$0.03 & 13$^{+7}_{-4}$ & 6.0$^{+0.6}_{-0.7}$ & -- & -- & -- \\
A143 & 58261.847 & 1141 & PL012 & 1.0$^{+0.7}_{-0.3}$ & 7$\pm$2 & 11$^{+1}_{-2}$ & 3.64$\pm$0.05 & 7$\pm$2 & 6.9$^{+0.8}_{-0.7}$ & 7.5$\pm$0.2 & 6$^{+11}_{-4}$ & 5$^{+2}_{-1}$ \\
A144 & 58262.504 & 764  & L012 & -- & 5$\pm$1 & 14$\pm$1 & 3.16$\pm$0.06 & 9$^{+6}_{-3}$ & 6.2$^{+0.8}_{-0.9}$ & 6.7$\pm$0.3 & 7$^{+11}_{-3}$ & 5$\pm$1 \\
A145 & 58263.330 & 1275 & L012 & -- & 4$_{-1}^{+2}$ & 13$\pm$2 & 3.04$\pm$0.05 & 7$\pm$2 & 7.0$^{+0.8}_{-0.9}$ & 5.8$^{+0.4}_{-0.6}$ & 2$^{+3}_{-1}$ & 8$^{+2}_{-3}$ \\
A146 & 58264.674 & 709 & L01 & -- & 7$\pm$1 & 16$\pm$1 & 2.77$\pm$0.04 & 12$^{+6}_{-3}$ & 7.2$^{+0.7}_{-0.8}$ & -- & -- & -- \\
A147 & 58265.962 & 873 & L012 & -- & 7$\pm$1 & 15$^{+1}_{-2}$ & 2.73$\pm$0.04 & 10$^{+4}_{-3}$ & 6.8$^{+0.7}_{-0.8}$ & 5.5$\pm$0.2 & 5$^{+5}_{-3}$ & 6$^{+2}_{-1}$ \\
A148 & 58266.669 & 862 & L012 & -- & 4$\pm$1 & 15$\pm$2 & 2.48$\pm$0.02 & 10$^{+4}_{-3}$ & 8.2$\pm$0.8 & 5.1$^{+0.2}_{-0.3}$ & 3$^{+3}_{-1}$ & 8$\pm$2 \\
A149 & 58268.664 & 994 & L012 & -- & 6.5$_{-0.7}^{+0.8}$ & 18$\pm$1 & 2.19$\pm$0.03 & 8$^{+3}_{-2}$ & 7.8$^{+0.7}_{-0.8}$ & 4.5$\pm$0.1 & 9$^{+7}_{-4}$ & 5$\pm$1 \\
A150 & 58269.500 & 1619 & L012 & -- & 5.6$_{-0.8}^{+0.7}$ & 17$\pm$1 & 2.22$\pm$0.03 & 7$^{+2}_{-1}$ & 8.2$\pm$0.6 & 4.4$\pm$0.1 & 6$^{+6}_{-3}$ & 6$^{+2}_{-1}$ \\
A152 & 58271.366 & 1787 & L012 & -- & 4.9$\pm$0.8 & 14$\pm$1& 3.11$\pm$0.03 & 8$\pm$2 & 7.1$\pm$0.6 & 6.2$\pm$0.2 & 4$^{+3}_{-2}$ & 6$\pm$1  \\
A153 & 58272.468 & 1452 & L012 & -- & 7.1$\pm$0.9 & 16$\pm$1 & 2.88$\pm$0.03 & 9$^{+3}_{-2}$ & 6.6$\pm$0.6 & 5.8$\pm$0.1 & 9$^{+11}_{-4}$ & 5$\pm$1 \\
A154 & 58273.432 & 1450 & L012 & -- & 7$\pm$1 & 16$\pm$1 & 2.94$\pm$0.02 & 13$^{+6}_{-4}$ & 5.8$\pm$0.6 & 5.8$^{+0.1}_{-0.2}$ & 8$^{+15}_{-5}$ & 4$^{+2}_{-1}$ \\
A155 & 58274.583 & 965 & L01 & -- & 6$_{-2}^{+1}$ & 15$^{+1}_{-2}$ & 2.78$\pm$0.03 & 12$^{+6}_{-3}$ & 6.6$\pm$0.7 & -- & -- & -- \\
A156 & 58275.484 & 2311 & L012 & -- & 5.4$\pm$0.8 & 15$\pm$1 & 2.65$\pm$0.02 & 8$\pm$2 & 7.5$^{+0.6}_{-0.5}$ & 5.2$\pm$0.1 & 5$^{+2}_{-1}$ & 7$\pm$1 \\
A157 & 58277.491 & 17109 & L012 & -- & 5.8$\pm$0.2 & 17.7$^{+0.3}_{-0.3}$ & 2.15$\pm$0.01 & 7.6$^{+0.6}_{-0.6}$ & 8.4$\pm$0.2 & 4.33$\pm$0.03 & 4.7$^{+0.7}_{-0.6}$ & 6.8$\pm$0.4 \\
A158 & 58277.994 & 5221 & L012 & -- & 5.9$\pm$0.4 & 17.4$^{+0.5}_{-0.6}$ & 2.18$\pm$0.01 & 7$\pm$1 & 8.5$^{+0.3}_{-0.4}$ & 4.41$^{+0.05}_{-0.06}$ & 5$^{+1}_{-1}$ & 6.6$\pm$0.6 \\
A159 & 58299.046 & 1127 & L012 & -- & 5.4$\pm$0.9 & 18$\pm$1 & 1.91$\pm$0.03 & 10$^{+4}_{-3}$ & 7.8$^{+0.8}_{-0.9}$ & 4.0$\pm$0.1 & 6$^{+4}_{-3}$ & 7$\pm$1 \\
\midrule
\multicolumn{3}{l}{\textbf{Linear decay phase}} \\
\midrule
& & & Model: & (P)owerlaw & \multicolumn{2}{l}{(L)orentzian (0)} & \multicolumn{3}{l}{Lorentzian (1) = QPO Main freq.} & \multicolumn{3}{l}{Lorentzian (2) = QPO 1st harmonic} \\
\nicer\/ & Date & Exp. & & $\Gamma$ & $\sigma_{0}$ & rms$_{0}$ & $\nu_{1}$ & Q$_{1}$ & rms$_{1}$ & $\nu_{2}$ & Q$_{2}$ & rms$_{2}$ \\
obs. & MJD & (s) & & & (Hz) & (\%) & (Hz) & & (\%) & (Hz) & & (\%) \\
\midrule
A160 & 58308.830 & 719 & L012 & -- & 6$\pm$1 & 18$^{+1}_{-2}$ & 1.92$^{+0.05}_{-0.02}$ & $>$9 & 7$\pm$1 & 3.90$\pm$0.09 & $>5$ & 6$^{+2}_{-2}$ \\
A161 & 58309.147 & 2705 & L012 & -- & 5.1$\pm$0.6 & 18.6$^{+0.8}_{-0.9}$ & 1.86$\pm$0.02 & 13$^{+4}_{-3}$ & 6.8$\pm$0.6 & 3.75$^{+0.05}_{-0.04}$ & 9$^{+5}_{-3}$ & 6$^{+1}_{-1}$ \\
A162 & 58310.839 & 812 & L01 & -- & 5.3$^{+0.9}_{-0.7}$ & 19$\pm$1 & 1.89$\pm$0.02 & 17$^{+15}_{-7}$ & 7$\pm$1 & -- & -- & -- \\
A163 & 58312.125 & 508 & L012 & -- & 4$^{+2}_{-1}$ & 15$\pm$3 & 1.87$^{+0.03}_{-0.04}$ & $>$6 & 7$^{+2}_{-1}$ & 4.3$\pm$0.3 & 2$^{+3}_{-1}$ & 11$^{+2}_{-3}$ \\
A164 & 58313.539 & 929 & L012 & -- & 3.9$^{+0.9}_{-0.8}$ & 17$^{+1}_{-2}$ & 1.74$\pm$0.02 & 11$^{+7}_{-4}$ & 8$\pm$1 & 3.67$\pm$0.07 & 9$^{+5}_{-3}$ & 7$^{+1}_{-1}$ \\
A165 & 58314.695 & 806 & L01 & -- & 6$\pm$1 & 20$\pm$1 & 2.00$^{+0.04}_{-0.06}$ & $>$6 & 7$\pm$1 & -- & -- & -- \\
A166 & 58315.595 & 812 & L012 & -- & 6$\pm$1 & 18$\pm$1 & 2.04$\pm$0.02 & $>$14 & 7$\pm$1 & 4.05$^{+0.06}_{-0.05}$ & $>11$ & 6$^{+1}_{-1}$ \\
A167 & 58316.560 & 537 & L01 & -- & 6$^{+2}_{-1}$ & 19$^{+1}_{-2}$ & 1.90$^{+0.06}_{-0.08}$ & 10$^{+13}_{-4}$ & 7$^{+1}_{-2}$ & -- & -- & -- \\
A168 & 58326.992 & 779 & L012 & -- & 4$\pm$1 & 17$\pm$2 & 1.48$^{+0.05}_{-0.04}$ & $>$9 & 8$^{+1}_{-2}$ & 3.04$\pm$0.09 & 7$^{+7}_{-3}$ & 7$^{+2}_{-2}$ \\
A169 & 58327.056 & 735 & L012 & -- & 1.3$^{+3.8}_{-0.5}$ & 13$\pm$2 & 1.47$\pm$0.03 & 6$^{+10}_{-2}$ & 10$^{+1}_{-3}$ & 3.0$^{+0.3}_{-0.6}$ & $>0.6$ & 14$^{+3}_{-9}$ \\
A170 & 58328.021 & 1561 & L012 & -- & 4.7$\pm$0.7 & 19$\pm$1 & 1.74$^{+0.03}_{-0.02}$ & $>$11 & 6.7$^{+0.8}_{-0.9}$ & 3.54$\pm$0.07 & 10$^{+8}_{-4}$ & 6$^{+1}_{-1}$ \\
A171 & 58329.051 & 1612 & L012 & -- & 2.9$^{+0.9}_{-0.7}$ & 16$^{+2}_{-1}$ & 1.86$\pm$0.02 & 13$^{+5}_{-3}$ & 8.1$^{+0.8}_{-0.9}$ & 3.8$^{+0.2}_{-0.1}$ & 4$^{+5}_{-2}$ & 8$^{+2}_{-2}$ \\
A172 & 58330.015 & 2974 & L012 & -- & 3$\pm$1 & 14$\pm$2 & 1.97$\pm$0.03 & 6$\pm$1 & 9$\pm$1 & 4.1$\pm$0.1 & 2$^{+2}_{-1}$ & 10$^{+2}_{-2}$ \\
A174 & 58332.663 & 558 & L01 & -- & 6$^{+2}_{-1}$ & 19$\pm$2 & 1.87$\pm$0.03 & 15$^{+10}_{-6}$ & 8$\pm$1 & -- & -- & -- \\
A175 & 58333.421 & 822 & L01 & -- & 7$\pm$1 & 20$\pm$1 & 1.73$^{+0.03}_{-0.04}$ & 10$^{+12}_{-4}$ & 7$\pm$1 & -- & -- & -- \\
A176 & 58334.333 & 795 & L012 & -- & 4$^{+1}_{-3}$ & 17$^{+2}_{-6}$ & 1.77$\pm$0.04 & 7$^{+6}_{-4}$ & 9$\pm$1 & 3.7$^{+0.2}_{-0.4}$ & 5$^{+11}_{-4}$ & 7$^{+2}_{-3}$ \\
A180 & 58343.136 & 928 & L01 & -- & 5.1$^{+0.8}_{-0.7}$ & 21$\pm$1 & 1.54$^{+0.03}_{-0.02}$ & $>$8 & 6$\pm$1 & -- & -- & -- \\
A181 & 58347.188 & 2543 & L012 & -- & 4.8$^{+0.8}_{-0.9}$ & 18$^{+1}_{-2}$ & 1.79$\pm$0.04 & 6$^{+2}_{-1}$ & 8.1$\pm$0.8 & 3.7$\pm$0.1 & 5$^{+3}_{-2}$ & 7$^{+2}_{-1}$ \\
A182 & 58350.534 & 1170 & L01 & -- & 6.0$^{+0.8}_{-0.7}$ & 22$\pm$1 & 1.62$^{+0.03}_{-0.02}$ & 15$^{+13}_{-7}$ & 7$\pm$1 & -- & -- & -- \\
A186 & 58358.279 & 1052 & L01 & -- & 6$\pm$1 & 21$\pm$1 & 1.88$^{+0.02}_{-0.03}$ & 12$^{+6}_{-4}$ & 8$\pm$1 & -- & -- & -- \\
A189 & 58361.304 & 1256 & L012 & -- & 5$\pm$1 & 18$^{+1}_{-2}$ & 1.77$\pm$0.02 & 15$^{+9}_{-5}$ & 7.8$^{+0.8}_{-0.9}$ & 3.6$\pm$0.1 & 8$^{+9}_{-4}$ & 7$^{+2}_{-2}$ \\
A190 & 58362.268 & 1261 & L012 & -- & 7$^{+1}_{-2}$ & 20$^{+1}_{-3}$ & 2.02$\pm$0.02 & 14$^{+9}_{-5}$ & 7.9$\pm$0.9 & 4.2$^{+0.4}_{-0.2}$ & $>$2 & 5$\pm$2 \\
A192 & 58365.357 & 1485 & L012 & -- & 5.9$\pm$0.9 & 20$\pm$1 & 2.02$\pm$0.02 & 13$^{+6}_{-4}$ & 8.7$\pm$0.8 & 4.2$\pm$0.1 & $>5$ & 5$^{+2}_{-2}$ \\
A193 & 58378.221 & 1287 & L012 & -- & 5.4$\pm$0.9 & 20$\pm$1 & 1.81$\pm$0.02 & 14$^{+11}_{-5}$ & 8$\pm$1 & 3.69 $\pm$0.07 & 13$^{+20}_{-8}$ & 6$^{+2}_{-1}$ \\
A195 & 58415.814 & 1526 & L012 & -- & 1.1$^{+0.4}_{-0.3}$ & 12$^{+1}_{-2}$ & 1.65$\pm$0.02 & 8$^{+4}_{-3}$ & 9$\pm$1 & 3.1$^{+0.3}_{-0.8}$ & 0.9$^{+0.8}_{-0.5}$ & 15$^{+3}_{-2}$ \\
A197 & 58420.314 & 3858 & L012 & -- & 4.6$^{+0.5}_{-0.4}$ & 20.7$^{+0.6}_{-0.7}$ & 1.25$\pm$0.01 & 14$^{+4}_{-3}$ & 7.2$\pm$0.6 &  2.5$\pm$0.1 & $>$20 & 3.2$^{+0.5}_{-0.6}$ \\
A198 & 58424.756 & 5647 & L01 & -- & 4.2$\pm$0.4 & 19.3$^{+0.7}_{-0.8}$ & 1.51$\pm$0.01 & 13$^{+5}_{-3}$ & 7.7$\pm$0.5 & 3.07$\pm$0.04 & 9$^{+5}_{-3}$ & 5.8$^{+0.9}_{-0.8}$ \\
A199 & 58425.013 & 6025 & L012 & -- & 4.2$\pm$0.4 & 19.3$^{+0.7}_{-0.8}$ & 1.62$^{+0.01}_{-0.02}$ & 9$^{+2}_{-1}$ & 7.9$\pm$0.5 & 3.35$^{+0.07}_{-0.05}$ & 7$^{+3}_{-2}$ & 6$^{+1}_{-1}$ \\
A201 & 58426.946 & 1720 & L012 & -- & 3$\pm$1 & 17$\pm$2 & 1.61$\pm$0.02 & 9$^{+4}_{-3}$ & 8$\pm$1 & 3.4$\pm$0.1 & 6$^{+5}_{-3}$ & 7$^{+2}_{-2}$ \\
A206 & 58438.095 & 636 & L012 & -- & 5$^{+1}_{-2}$ & 20$^{+2}_{-4}$ & 1.49$^{+0.06}_{-0.05}$ & 7$^{+13}_{-4}$ & 8$^{+1}_{-2}$ & 3.3$^{+0.1}_{-0.3}$ & $>$2 & 6$^{+2}_{-3}$ \\
A208 & 58440.027 & 856 & L01 & -- & 4.7$^{+0.9}_{-0.8}$ & 21$\pm$1 & 1.39$\pm$0.03 & $>$6 & 6$^{+1}_{-2}$ & -- & -- & -- \\
A209 & 58449.092 & 3065 & L012 & -- & 1.3$^{+1.0}_{-0.4}$ & 12$^{+1}_{-2}$ & 1.39$\pm$0.04 & 4$^{+3}_{-1}$ & 9$^{+2}_{-4}$ & 2.8$^{+0.2}_{-0.4}$ & 1.3$^{+1.0}_{-0.6}$ & 13$\pm$2 \\
A210 & 58450.958 & 804 & L01 & -- & 4$\pm$1 & 21$\pm$1 & 1.28$^{+0.02}_{-0.03}$ & $>$9 & 7$^{+1}_{-2}$ & -- & -- & -- \\
A211 & 58451.022 & 2391 & L012 & -- & 0.6$^{+0.4}_{-0.2}$ & 9$\pm$2 & 1.55$^{+0.04}_{-0.49}$ & $>$10 & 5$^{+1}_{-2}$ & 3.2$^{+0.3}_{-0.6}$ & 1.1$^{+0.8}_{-0.5}$ & 13$\pm$2 \\
A212 & 58452.888 & 798 & L01 & -- & 5$\pm$1 & 19$\pm$2 & 1.57$\pm$0.05 & 10$^{+7}_{-4}$ & 7$\pm$1 & -- & -- & -- \\
A213 & 58453.725 & 904 & L01 & -- & 4$\pm$1 & 20$^{+1}_{-2}$ & 1.45$\pm$0.04 & 8$^{+7}_{-3}$ & 8$^{+1}_{-2}$ & -- & -- & -- \\
A214 & 58455.205 & 850 & L01 & -- & 5$\pm$1 & 21$^{+1}_{-2}$ & 1.36$^{+0.05}_{-0.04}$ & 7$^{+7}_{-3}$ & 8$^{+1}_{-2}$ & -- & -- & -- \\
A215 & 58456.234 & 2826 & L012 & -- & 1.2$^{+0.6}_{-0.3}$ & 13$^{+2}_{-1}$ & 1.35$\pm$0.02 & 7$\pm$2 & 9$\pm$1 & 2.8$^{+0.1}_{-0.2}$ & 1.5$^{+1.0}_{-0.6}$ & 13$\pm$2 \\
A216 & 58457.264 & 1278 & L012 & -- & 4$\pm$1 & 20$\pm$1 & 1.37$^{+0.02}_{-0.01}$ & $>$15 & 7$\pm$1 & 2.9$\pm$0.1 & $>$10 & 5$^{+1}_{-2}$ \\
A217 & 58458.229 & 1185 & L012 & -- & 4$\pm$1 & 20$^{+1}_{-2}$ & 1.35$\pm$0.02 & $>$9 & 6$\pm$1 & 2.8$\pm$0.1 & $>$6 & 6$\pm$2 \\
A218 & 58459.259 & 1262 & L012 & -- & 3$\pm$1 & 18$^{+2}_{-4}$ & 1.16$^{+0.03}_{-0.02}$ & 7$^{+4}_{-2}$ & 9$^{+2}_{-1}$ & 2.4$\pm$0.1 & 6$^{+8}_{-4}$ & $>$5 \\
A220 & 58461.255 & 853 & L01 & -- & 5$\pm$1 & 21$^{+1}_{-2}$ & 1.35$^{+0.02}_{-0.03}$ & $>$8 & 7$\pm$1 & -- & -- & --  \\
A221 & 58462.092 & 764 & L01 & -- & 6$\pm$1 & 21$\pm$2 & 1.60$\pm$0.05 & 8$^{+6}_{-3}$ & 8$^{+1}_{-2}$ & -- & -- & --  \\
A222 & 58463.251 & 510 & L01 & -- & 6$\pm$1 & 21$\pm$2 & 1.60$\pm$0.05 & 8$^{+6}_{-3}$ & 8$^{+1}_{-2}$ & -- & -- & --  \\
\midrule
\multicolumn{3}{l}{\textbf{Rebrightening phase}} \\
\midrule
& & & Model: & (P)owerlaw & \multicolumn{2}{l}{(L)orentzian (0)} & \multicolumn{3}{l}{Lorentzian (1) = QPO Main freq.} & \multicolumn{3}{l}{Lorentzian (2) = QPO 1st harmonic} \\
\nicer\/ & Date & Exp. & & $\Gamma$ & $\sigma_{0}$ & rms$_{0}$ & $\nu_{1}$ & Q$_{1}$ & rms$_{1}$ & $\nu_{2}$ & Q$_{2}$ & rms$_{2}$ \\
obs. & MJD & (s) & & & (Hz) & (\%) & (Hz) & & (\%) & (Hz) & & (\%) \\
\midrule
B101 & 58548.692 & 3818 & PL012 & 0.7$^{+0.4}_{-0.2}$ & 5$\pm$1 & 16$\pm$2 & 2.07$\pm$0.01 & 13$^{+5}_{-4}$ & 6.6$\pm$0.6 & 4.20$^{+0.07}_{-0.06}$ & 9$^{+5}_{-3}$ & 6$\pm$1 \\
B201 & 58555.709 & 3107 & PL012 & 1.0$^{+1.4}_{-0.4}$ & 6$\pm$1 & 18$^{+1}_{-3}$ & 2.03$\pm$0.01 & 18$^{+9}_{-5}$ & 6.8$\pm$0.6 & 4.06$\pm$0.07 & 9$^{+5}_{-3}$ & 6$\pm$1 \\
B301 & 58562.664 & 2867 & PL012 & 0.6$^{+0.2}_{-0.1}$ & 4$\pm$1 & 13$\pm$2 & 2.23$\pm$0.02 & 14$^{+8}_{-4}$ & 6.8$\pm$0.7 & 4.55$^{+0.05}_{-0.06}$ & 9$^{+5}_{-3}$ & 7$\pm$1 \\
B401 & 58569.429 & 3222 & PL012 & 1.1$^{+0.4}_{-0.2}$ & 5$\pm$1 & 16$^{+1}_{-2}$ & 2.42$\pm$ 0.02 & 11$^{+4}_{-3}$ & 6.6$\pm$ 0.7 & 4.94$\pm$0.08 & 9$^{+5}_{-3}$ & 6$\pm$1 \\
B501 & 58577.106 & 731 & PL012 & 1.5$^{+0.5}_{-0.4}$ & 6$^{+4}_{-2}$ & 16$^{+3}_{-4}$ & 2.42$^{+0.06}_{-0.07}$ & 9$^{+11}_{-5}$ & 7$\pm$2 & 4.9$^{+0.3}_{-0.2}$ & $>$3 & 7$\pm$2 \\
B601 & 58583.037 & 3011 & PL012 & 1.0$^{+0.3}_{-0.2}$ & 5$\pm$1 & 16$^{+2}_{-3}$ & 2.21$\pm$0.01 & 15$^{+6}_{-3}$ & 8.0$\pm$0.6 & 4.49$^{+0.09}_{-0.08}$ & 7$^{+4}_{-3}$ & 7$\pm$1 \\
B701 & 58589.157 & 2983 & PL012 & 1.2$^{+0.2}_{-0.2}$ & 6$\pm$2 & 15$\pm$2 & 2.49$\pm$0.03 & 8$\pm$2 & 7.4$\pm$0.8 & 5.17$\pm$0.14 & 4$^{+2}_{-1}$ & 8$^{+2}_{-1}$ \\
B801 & 58596.557 & 3766 & PL012 & 1.3$\pm$0.1 & 6$\pm$1 & 17$\pm$2 & 2.08$\pm$0.02 & 8$^{+2}_{-1}$ & 10.2$\pm$0.7 & 4.30$^{+0.08}_{-0.07}$ & 7$^{+4}_{-2}$ & 7$\pm$1 \\
B901 & 58604.414 & 3601 & PL012 & 1.4$\pm$0.1 & 7$\pm$3 & 13$^{+3}_{-4}$ & 3.10$\pm$0.05 & 5$^{+2}_{-1}$ & 9$\pm$1 & 6.2$^{+0.2}_{-0.3}$ & 5$^{+5}_{-2}$ & 7$\pm$2 \\
C001 & 58608.537 & 4432 & PL1 & 1.32$\pm$0.05 & -- & -- & 3.2$\pm$0.2 & 3$^{+3}_{-2}$ & 0.09$\pm$0.02 & -- & -- & -- \\
\midrule
\multicolumn{3}{l}{\textbf{Obscured phase}} \\
\midrule
& & & Model: & (P)owerlaw & \multicolumn{2}{l}{(L)orentzian (0)} & \multicolumn{3}{l}{Lorentzian (1) = QPO Main freq.} & \multicolumn{3}{l}{Lorentzian (2) = QPO 1st harmonic} \\
\nicer\/ & Date & Exp. & & $\Gamma$ & $\sigma_{0}$ & rms$_{0}$ & $\nu_{1}$ & Q$_{1}$ & rms$_{1}$ & $\nu_{2}$ & Q$_{2}$ & rms$_{2}$ \\
obs. & MJD & (s) & & & (Hz) & (\%) & (Hz) & & (\%) & (Hz) & & (\%) \\
\midrule
C202 & 58622.518 & 14421 & P & 1.3$\pm$0.2 & -- & -- & -- & -- & -- & -- & -- & -- \\
C203 & 58623.033 & 4618 & P & 1.2$^{+0.4}_{-0.3}$ & -- & -- & -- & -- & -- & -- & -- & -- \\
C204 & 58624.065 & 6830 & P & 1.9$\pm$0.1 & -- & -- & -- & -- & -- & -- & -- & -- \\
C205 & 58625.030 & 4314 & P & 2.0$\pm$0.1 & -- & -- & -- & -- & -- & -- & -- & --  \\
C206 & 58625.998 & 3673 & P & 1.7$\pm$0.2 & -- & -- & -- & -- & -- & -- & -- & -- \\
C207 & 58627.417 & 6401 & P & 1.8$^{+0.3}_{-0.2}$ & -- & -- & -- & -- & -- & -- & -- & -- \\
C208 & 58628.057 & 4627 & P & 2.3$^{+0.3}_{-0.2}$ & -- & -- & -- & -- & -- & -- & -- & -- \\
C209 & 58629.024 & 1532 & -- & -- & -- & -- & -- & -- & -- & -- & -- & -- \\
C301 & 58631.089 & 15866 & P & 1.7$^{+0.5}_{-0.4}$ & -- & -- & -- & -- & -- & -- & -- & -- \\ 
C302 & 58632.247 & 8146 & P & 2.0$^{+0.4}_{-0.3}$ & -- & -- & -- & -- & -- & -- & -- & -- \\
C303 & 58633.601 & 4790 & P & 1.7$\pm$0.3 & -- & -- & -- & -- & -- & -- & -- & -- \\
C304 & 58634.246 & 3514 & -- & -- & -- & -- & -- & -- & -- & -- & -- & -- \\
\bottomrule
\end{longtable}
}
}
\longtab[1]{
{\small\tabcolsep=3pt  
\begin{longtable}[1]{lllllllllllllll}
\caption{\nicer\/ spectral analysis parameters.} \\
\toprule
\multicolumn{3}{l}{\textbf{Exponential decay phase}} \\
\midrule
\nicer\/ & N(H) & N(Mg) & N(Al) & N(Si) & N(S) & N(Ca) & $\Gamma$ & log $\xi$ & R$_{\mathrm{f}}$ & R$_{\mathrm{in}}$ & F$_{\mathrm{abs}}$ & F$_{\mathrm{unabs}}$ & $\chi^{2}_{\mathrm{red}}$ \\
obs. & \multicolumn{6}{c}{10$^{22}$ atoms/cm$^{2}$} & & & & & 10$^{-8}$ erg/s/cm$^{2}$ & 10$^{-8}$ erg/s/cm$^{2}$ & \\
\midrule
A137 & 2.40$\pm$0.01 & 3.7$\pm$0.2 & 17$\pm$1 & 3.2$\pm$0.1 & 2.2$\pm$0.2 & 4$\pm$1 & 2.71$\pm$0.02 & 4.70$\pm$0.05 & 0.17$^{+0.03}_{-0.03}$ & 44$^{+54}_{-21}$ & 0.58$\pm$0.04 & 2.02$\pm$0.13 & 1.32 \\
A138 & 2.38$\pm$0.01 & 3.2$\pm$0.3 & 16$\pm$2 & 2.9$\pm$0.1 & 2.1$\pm$0.4 & 5$\pm$1 & 2.62$^{+0.02}_{-0.03}$ & 4.48$^{+0.07}_{-0.10}$ & 0.17 & 44 & 0.56$\pm$0.02 & 1.82$\pm$0.07 & 1.06 \\ 
A139 & 2.30$\pm$0.01 & 3.5$\pm$0.3 & 10$\pm$2 & 2.8$\pm$0.2 & 1.7$\pm$0.4 & 8$\pm$2 & 2.45$^{+0.04}_{-0.03}$ & 4.12$^{+0.06}_{-0.09}$ & 0.17 & 44 & 0.53$\pm$0.03 & 1.48$\pm$0.07 & 1.43 \\ 
A140 & 2.30$\pm$0.01 & 3.1$\pm$0.4 & 11$\pm$2 & 2.7$\pm$0.2 & 2.2$\pm$0.5 & 8$\pm$2 & 2.44$\pm$0.04 & 4.04$^{+0.05}_{-0.06}$ & 0.17 & 44 & 0.52$\pm$0.03 & 1.44$\pm$0.07 & 1.20 \\ 
A142 & 2.29$\pm$0.02 & 3.2$^{+0.5}_{-0.4}$ & 11$\pm$3 & 2.9$\pm$0.2 & 2.7$\pm$0.6 & 11$^{+2.5}_{-2.4}$ & 2.43$\pm$0.04 & 4.04$^{+0.08}_{-0.07}$ & 0.17 & 44 & 0.40$\pm$0.03 & 1.13$\pm$0.06 & 1.19 \\ 
A143 & 2.32$\pm$0.02 & 3.3$\pm$0.4 & 14$\pm$3 & 3.0$\pm$0.2 & 2.4$^{+0.52}_{-0.58}$ & 11$\pm$2 & 2.54$^{+0.04}_{-0.03}$ & 4.3$\pm$0.1 & 0.17 & 44 & 0.44$\pm$0.03 & 1.34$\pm$0.07 & 1.07 \\ 
A144 & 2.29$\pm$0.02 & 2.7$\pm$0.5 & 9$\pm$3 & 3.0$^{+0.2}_{-0.3}$ & 2.3$\pm$0.7 & 11$\pm$3 & 2.40$^{+0.04}_{-0.05}$ & 4.0$\pm$0.1 & 0.17 & 44 & 0.42$\pm$0.03 & 1.14$\pm$0.06 & 1.04 \\ 
A145 & 2.28$^{+0.02}_{-0.01}$ & 2.5$\pm$0.4 & 11$\pm$2 & 2.8$\pm$0.2 & 2.7$^{+0.60}_{-0.54}$ & 10$\pm$2 & 2.38$\pm$0.03 & 4.0$\pm$0.1 & 0.17 & 44 & 0.40$\pm$0.02 & 1.08$\pm$0.04 & 1.02 \\ 
A146 & 2.25$\pm$0.02 & 2.8$^{+0.5}_{-0.6}$ & 8$^{+3}_{-4}$ & 2.9$\pm$0.3 & 2.6$\pm$0.8 & 11$\pm$3 & 2.31$\pm$0.05 & 3.7$\pm$0.1 & 0.17 & 44 & 0.38$\pm$0.02 & 0.97$\pm$0.05 & 1.28 \\ 
A147 & 2.24$\pm$0.02 & 2.9$\pm$0.5 & 11$\pm$3 & 2.8$^{+0.3}_{-0.2}$ & 3.0$\pm$0.7 & 14$\pm$3 & 2.32$\pm$0.04 & 3.8$\pm$0.1 & 0.17 & 44 & 0.38$\pm$0.02 & 0.99$\pm$0.04 & 1.01 \\ 
A148 & 2.26$\pm$0.02 & 2.1$\pm$0.5 & 9$\pm$3 & 2.8$\pm$0.3 & 3.1$^{+0.75}_{-0.74}$ & 11$\pm$3 & 2.25$\pm$0.05 & 3.6$\pm$0.1 & 0.17 & 44 & 0.37$\pm$0.02 & 0.92$\pm$0.05 & 1.24 \\ 
A149 & 2.21$\pm$0.02 & 2.6$\pm$0.5 & 9$\pm$3 & 3.0$\pm$0.2 & 3.7$^{+0.65}_{-0.57}$ & 13$\pm$3 & 2.18$^{+0.04}_{-0.03}$ & 3.6$\pm$0.1 & 0.17 & 44 & 0.35$\pm$0.01 & 0.84$\pm$0.02 & 1.33 \\ 
A150 & 2.22$\pm$0.01 & 2.7$\pm$0.4 & 12$\pm$2 & 3.1$\pm$0.2 & 4.3$^{+1.03}_{-0.56}$ & 13$\pm$2 & 2.21$^{+0.03}_{-0.04}$ & 3.6$\pm$0.1 & 0.17 & 44 & 0.35$\pm$0.01 & 0.86$\pm$0.03 & 1.14 \\ 
A152 & 2.26$\pm$0.01 & 2.8$\pm$0.3 & 10$\pm$2 & 2.8$\pm$0.2 & 2.3$^{+0.44}_{-0.46}$ & 10$\pm$2 & 2.34$^{+0.03}_{-0.02}$ & 4.00$^{+0.07}_{-0.06}$ & 0.17 & 44 & 0.37$\pm$0.01 & 0.97$\pm$0.03 & 1.19 \\ 
A153 & 2.26$^{+0.02}_{-0.01}$ & 2.9$\pm$0.4 & 12$\pm$2 & 3.0$\pm$0.2 & 3.1$^{+0.55}_{-0.54}$ & 11$\pm$2 & 2.37$\pm$0.03 & 3.79$^{+0.06}_{-0.07}$ & 0.17 & 44 & 0.35$\pm$0.01 & 0.94$\pm$0.03 & 1.20 \\ 
A154 & 2.25$\pm$0.01 & 3.1$\pm$0.4 & 10$\pm$2 & 3.0$\pm$0.2 & 3.0$\pm$0.6 & 13$\pm$2 & 2.37$\pm$0.03 & 3.86$^{+0.07}_{-0.09}$ & 0.17 & 44 & 0.34$\pm$0.01 & 0.93$\pm$0.03 & 1.29 \\  
A155 & 2.25$\pm$0.02 & 3.0$\pm$0.5 & 13$\pm$3 & 3.0$\pm$0.2 & 3.8$^{+0.65}_{-0.64}$ & 14$\pm$2 & 2.38$^{+0.03}_{-0.04}$ & 3.9$\pm$0.1 & 0.17 & 44 & 0.34$\pm$0.02 & 0.95$\pm$0.04 & 1.09 \\ 
A156 & 2.26$\pm$0.01 & 2.1$\pm$0.3 & 7$\pm$2 & 2.7$\pm$0.2 & 2.3$^{+0.42}_{-0.51}$ & 10$\pm$2 & 2.24$\pm$0.03 & 3.70$^{+0.05}_{-0.06}$ & 0.17 & 44 & 0.34$\pm$0.01 & 0.81$\pm$0.02 & 1.15 \\ 
A157 & 2.19$\pm$0.01 & 2.7$\pm$0.1 & 7$\pm$1 & 3.2$\pm$0.1 & 4.1$\pm$0.4 & 10$\pm$2 & 2.12$\pm$0.02 & 3.70$^{+0.05}_{-0.07}$ & 0.23$^{+0.12}_{-0.04}$ & 100$^{+0}_{-62}$ & 0.32$\pm$0.03 & 0.74$\pm$0.08 & 1.22 \\ 
A158 & 2.19$\pm$0.01 & 2.8$\pm$0.2 & 7$\pm$2 & 3.2$\pm$0.2 & 4.3$\pm$0.5 & 6$\pm$2 & 2.10$\pm$0.02 & 3.71$^{+0.05}_{-0.04}$ & 0.35 & 100 & 0.31$\pm$0.01 & 0.71$\pm$0.01 & 1.15 \\ 
A159 & 2.20$\pm$0.01 & =N(H) & 3.4$\pm$0.3 & =N(Al) & =N(Al) & =N(Al) & 2.00$\pm$0.03 & 3.50$^{+0.06}_{-0.08}$ & 0.35 & 100 & 0.24$\pm$0.01 & 0.51$\pm$0.01 & 1.01 \\ 
\midrule
\multicolumn{3}{l}{\textbf{Linear decay phase}} \\
\midrule
\nicer\/ & N(H) & N(Mg) & N(Al) & N(Si) & N(S) & N(Ca) & $\Gamma$ & log $\xi$ & R$_{\mathrm{f}}$ & R$_{\mathrm{in}}$ & F$_{\mathrm{abs}}$ & F$_{\mathrm{unabs}}$ & $\chi^{2}_{\mathrm{red}}$ \\
obs. & \multicolumn{6}{c}{10$^{22}$ atoms/cm$^{2}$} & & & & & 10$^{-8}$ erg/s/cm$^{2}$ & 10$^{-8}$ erg/s/cm$^{2}$ & \\
\midrule
A160 & 2.21$^{+0.01}_{-0.02}$ & =N(H) & 2.8$\pm$0.3 & =N(Al) & =N(Al) & =N(Al) & 2.02$^{+0.03}_{-0.04}$ & 3.6$\pm$0.1 & 0.35 & 100 & 0.212$\pm$0.004 & 0.444$\pm$0.006 & 1.09 \\
A161 & 2.20$\pm$0.01 & =N(H) & 3.2$\pm$0.2 & =N(Al) & =N(Al) & =N(Al) & 2.03$\pm$0.02 & 3.54$^{+0.05}_{-0.06}$ & 0.35 & 100 & 0.204$\pm$0.002 & 0.436$\pm$0.004 & 1.12 \\
A162 & 2.20$^{+0.01}_{-0.02}$ & =N(H) & 3.1$^{+0.3}_{-0.4}$ & =N(Al) & =N(Al) & =N(Al) & 2.02$\pm$0.04 & 3.6$\pm$0.1 & 0.35 & 100 & 0.204$\pm$0.004 & 0.432$\pm$0.007 & 1.09 \\
A163 & 2.22$\pm$0.02 & =N(H) & 2.9$^{+0.4}_{-0.3}$ & =N(Al) & =N(Al) & =N(Al) & 2.05$\pm$0.05 & 3.5$\pm$0.1 & 0.35 & 100 & 0.196$\pm$0.006 & 0.421$\pm$0.009 & 1.07 \\
A164 & 2.20$^{+0.01}_{-0.02}$ & =N(H) & 2.9$^{+0.2}_{-0.3}$ & =N(Al) & =N(Al) & =N(Al) & 2.01$^{+0.03}_{-0.04}$ & 3.43$^{+0.05}_{-0.07}$ & 0.35 & 100 & 0.194$\pm$0.005 & 0.407$\pm$0.009 & 1.11 \\
A165 & 2.21$\pm$0.02 & =N(H) & 3.0$^{+0.4}_{-0.8}$ & =N(Al) & =N(Al) & =N(Al) & 2.04$\pm$0.04 & 3.60$^{+0.10}_{-0.15}$ & 0.35 & 100 & 0.200$\pm$0.006 & 0.430$\pm$0.011 & 1.08 \\
A166 & 2.21$^{+0.02}_{-0.01}$ & =N(H) & 3.4$^{+0.3}_{-0.4}$ & =N(Al) & =N(Al) & =N(Al) & 2.08$\pm$0.04 & 3.6$\pm$0.1 & 0.35 & 100 & 0.196$\pm$0.005 & 0.433$\pm$0.008 & 0.98 \\
A167 & 2.18$\pm$0.02 & =N(H) & 3.2$\pm$0.5 & =N(Al) & =N(Al) & =N(Al) & 1.97$^{+0.04}_{-0.05}$ & 3.6$^{+0.1}_{-0.2}$ & 0.35 & 100 & 0.185$\pm$0.002 & 0.383$\pm$0.005 & 1.18 \\
A168 & 2.22$\pm$0.02 & =N(H) & 3.5$^{+0.6}_{-0.4}$ & =N(Al) & =N(Al) & =N(Al) & 1.98$\pm$0.04 & 3.6$^{+0.1}_{-0.2}$ & 0.35 & 100 & 0.164$\pm$0.003 & 0.347$\pm$0.005 & 1.05 \\
A169 & 2.20$\pm$0.02 & =N(H) & 3.1$^{+0.3}_{-0.6}$ & =N(Al) & =N(Al) & =N(Al) & 1.93$\pm$0.04 & 3.8$^{+0.2}_{-0.3}$ & 0.35 & 100 & 0.169$\pm$0.007 & 0.343$\pm$0.014 & 0.98 \\
A170 & 2.21$\pm$0.01 & =N(H) & 3.1$\pm$0.3 & =N(Al) & =N(Al) & =N(Al) & 1.98$\pm$0.03 & 3.51$^{+0.06}_{-0.09}$ & 0.35 & 100 & 0.172$\pm$0.002 & 0.359$\pm$0.004 & 1.03 \\
A171 & 2.19$\pm$0.01 & =N(H) & 3.4$\pm$0.3 & =N(Al) & =N(Al) & =N(Al) & 2.00$\pm$0.03 & 3.52$^{+0.06}_{-0.08}$ & 0.35 & 100 & 0.173$\pm$0.002 & 0.366$\pm$0.003 & 1.09 \\
A172 & 2.19$\pm$0.01 & =N(H) & 3.1$\pm$0.2 & =N(Al) & =N(Al) & =N(Al) & 1.99$\pm$0.02 & 3.52$^{+0.05}_{-0.06}$ & 0.35 & 100 & 0.175$\pm$0.001 & 0.365$\pm$0.002 & 1.12 \\
A174 & 2.19$\pm$0.02 & =N(H) & 3.3$\pm$0.5 & =N(Al) & =N(Al) & =N(Al) & 2.00$^{+0.04}_{-0.05}$ & 3.5$^{+0.1}_{-0.2}$ & 0.35 & 100 & 0.165$\pm$0.003 & 0.350$\pm$0.005 & 1.01 \\
A175 & 2.20$\pm$0.01 & =N(H) & 2.8$\pm$0.3 & =N(Al) & =N(Al) & =N(Al) & 1.97$\pm$0.04 & 3.5$\pm$0.1 & 0.35 & 100 & 0.168$\pm$0.004 & 0.343$\pm$0.008 & 1.13 \\
A176 & 2.18$\pm$0.02 & =N(H) & 3.3$\pm$0.4 & =N(Al) & =N(Al) & =N(Al) & 1.96$\pm$0.04 & 3.6$^{+0.1}_{-0.2}$ & 0.35 & 100 & 0.165$\pm$0.002 & 0.340$\pm$0.004 & 1.05 \\
A180 & 2.18$\pm$0.01 & =N(H) & 3.1$\pm$0.3 & =N(Al) & =N(Al) & =N(Al) & 1.91$\pm$0.04 & 3.4$\pm$0.1 & 0.35 & 100 & 0.164$\pm$0.004 & 0.328$\pm$0.009 & 1.16 \\
A181 & 2.17$\pm$0.01 & =N(H) & 3.0$\pm$0.2 & =N(Al) & =N(Al) & =N(Al) & 1.89$\pm$0.02 & 3.6$\pm$0.1 & 0.35 & 100 & 0.165$\pm$0.001 & 0.325$\pm$0.003 & 1.12 \\
A182 & 2.17$\pm$0.02 & =N(H) & 2.9$\pm$0.4 & =N(Al) & =N(Al) & =N(Al) & 1.88$\pm$0.04 & 3.5$^{+0.1}_{-0.2}$ & 0.35 & 100 & 0.157$\pm$0.002 & 0.306$\pm$0.005 & 1.19 \\
A186 & 2.16$\pm$0.02 & =N(H) & 3.0$\pm$0.4 & =N(Al) & =N(Al) & =N(Al) & 1.86$\pm$0.04 & 3.5$^{+0.1}_{-0.2}$ & 0.35 & 100 & 0.161$\pm$0.002 & 0.315$\pm$0.005 & 1.13 \\
A189 & 2.19$\pm$0.01 & =N(H) & 3.4$\pm$0.3 & =N(Al) & =N(Al) & =N(Al) & 1.97$\pm$0.03 & 3.6$\pm$0.1 & 0.35 & 100 & 0.151$\pm$0.001 & 0.314$\pm$0.004 & 1.06 \\
A190 & 2.20$\pm$0.01 & =N(H) & 3.3$\pm$0.4 & =N(Al) & =N(Al) & =N(Al) & 2.00$^{+0.03}_{-0.04}$ & 3.6$\pm$0.1 & 0.35 & 100 & 0.158$\pm$0.002 & 0.335$\pm$0.003 & 1.08\\
A192 & 2.18$\pm$0.01 & =N(H) & 3.0$\pm$0.3 & =N(Al) & =N(Al) & =N(Al) & 1.98$\pm$0.03 & 3.6$\pm$0.1 & 0.35 & 100 & 0.157$\pm$0.002 & 0.324$\pm$0.004 & 1.02 \\
A193 & 2.19$\pm$0.01 & =N(H) & 3.2$^{+0.3}_{-0.4}$ & =N(Al) & =N(Al) & =N(Al) & 1.98$\pm$0.03 & 3.47$^{+0.06}_{-0.09}$ & 0.35 & 100 & 0.150$\pm$0.002 & 0.312$\pm$0.005 & 0.96 \\
A195 & 2.20$\pm$0.01 & =N(H) & 3.4$^{+0.2}_{-0.3}$ & =N(Al) & =N(Al) & =N(Al) & 2.01$^{+0.03}_{-0.04}$ & 3.47$^{+0.06}_{-0.08}$ & 0.35 & 100 & 0.131$\pm$0.002 & 0.279$\pm$0.004 & 1.02 \\
A197 & 2.15$^{+0.01}_{-0.02}$ & 3.2$\pm$0.4 & 14$\pm$3 & 3.7$\pm$0.2 & 4.3$\pm$0.5 & 12$\pm$2 & 2.00$\pm$0.03 & 3.59$^{+0.07}_{-0.09}$ & 0.35 & 100 & 0.127$\pm$0.002 & 0.282$\pm$0.003 & 1.17 \\
A198 & 2.17$\pm$0.01 & 2.9$\pm$0.3 & 12$\pm$2 & 3.4$\pm$0.2 & 4.9$\pm$0.6 & 7$^{+4}_{-3}$ & 2.00$^{+0.03}_{-0.04}$ & 3.52$^{+0.05}_{-0.07}$ & 0.35 & 100 & 0.131$\pm$0.002 & 0.285$\pm$0.003 & 1.03 \\
A199 & 2.17$\pm$0.01 & 2.8$\pm$0.3 & 8$\pm$2 & 3.1$\pm$0.2 & 3.0$\pm$0.4 & 8$\pm$2 & 1.97$\pm$0.02 & 3.62$^{+0.06}_{-0.08}$ & 0.35 & 100 & 0.135$\pm$0.002 & 0.283$\pm$0.004 & 1.22 \\
A201 & 2.14$\pm$0.01 & =N(H) & 3.1$\pm$0.3 & =N(Al) & =N(Al) & =N(Al) & 1.81$\pm$0.03 & 3.5$\pm$0.1 & 0.35 & 100 & 0.138$\pm$0.003 & 0.262$\pm$0.005 & 1.22 \\
A206 & 2.20$\pm$0.02 & =N(H) & 3.1$\pm$0.4 & =N(Al) & =N(Al) & =N(Al) & 1.93$\pm$0.05 & 3.4$\pm$0.1 & 0.35 & 100 & 0.132$\pm$0.005 & 0.266$\pm$0.010 & 1.16 \\
A208 & 2.21$\pm$0.02 & =N(H) & 3.1$\pm$0.3 & =N(Al) & =N(Al) & =N(Al) & 1.93$\pm$0.04 & 3.4$\pm$0.1 & 0.35 & 100 & 0.131$\pm$0.004 & 0.264$\pm$0.008 & 1.01 \\
A209 & 2.19$\pm$0.02 & =N(H) & 3.3$\pm$0.3 & =N(Al) & =N(Al) & =N(Al) & 1.89$\pm$0.04 & 3.5$\pm$0.1 & 0.35 & 100 & 0.124$\pm$0.003 & 0.246$\pm$0.007 & 1.25 \\
A210 & 2.21$\pm$0.02 & =N(H) & 3.0$\pm$0.4 & =N(Al) & =N(Al) & =N(Al) & 1.90$\pm$0.05 & 3.4$\pm$0.1 & 0.35 & 100 & 0.122$\pm$0.004 & 0.244$\pm$0.009 & 1.19 \\
A211 & 2.21$\pm$0.01 & =N(H) & 3.0$\pm$0.2 & =N(Al) & =N(Al) & =N(Al) & 1.92$\pm$0.03 & 3.6$\pm$0.1 & 0.35 & 100 & 0.126$\pm$0.001 & 0.252$\pm$0.002 & 1.13 \\
A212 & 2.21$\pm$0.02 & =N(H) & 2.9$\pm$0.4 & =N(Al) & =N(Al) & =N(Al) & 1.92$\pm$0.05 & 3.6$^{+0.1}_{-0.2}$ & 0.35 & 100 & 0.124$\pm$0.002 & 0.249$\pm$0.005 & 1.20 \\
A213 & 2.21$\pm$0.02 & =N(H) & 3.4$\pm$0.3 & =N(Al) & =N(Al) & =N(Al) & 1.94$\pm$0.04 & 3.5$^{+0.1}_{-0.2}$ & 0.35 & 100 & 0.117$\pm$0.002 & 0.241$\pm$0.004 & 1.14 \\
A214 & 2.21$\pm$0.02 & =N(H) & 3.0$\pm$0.3 & =N(Al) & =N(Al) & =N(Al) & 1.91$^{+0.05}_{-0.04}$ & 3.7$^{+0.1}_{-0.2}$ & 0.35 & 100 & 0.122$\pm$0.004 & 0.245$\pm$0.009 & 1.16 \\
A215 & 2.21$\pm$0.01 & =N(H) & 2.9$\pm$0.2 & =N(Al) & =N(Al) & =N(Al) & 1.87$\pm$0.02 & 3.45$^{+0.05}_{-0.07}$ & 0.35 & 100 & 0.124$\pm$0.002 & 0.243$\pm$0.004 & 1.26 \\
A216 & 2.19$\pm$0.02 & =N(H) & 2.9$\pm$0.3 & =N(Al) & =N(Al) & =N(Al) & 1.85$\pm$0.04 & 3.5$\pm$0.1 & 0.35 & 100 & 0.123$\pm$0.002 & 0.239$\pm$0.005 & 1.22 \\
A217 & 2.21$\pm$0.02 & =N(H) & 2.6$\pm$0.3 & =N(Al) & =N(Al) & =N(Al) & 1.85$\pm$0.04 & 3.5$^{+0.1}_{-0.2}$ & 0.35 & 100 & 0.124$\pm$0.003 & 0.239$\pm$0.006 & 1.16 \\
A218 & 2.20$\pm$0.02 & =N(H) & 3.0$\pm$0.3 & =N(Al) & =N(Al) & =N(Al) & 1.82$\pm$0.04 & 3.7$\pm$0.2 & 0.35 & 100 & 0.120$\pm$0.004 & 0.230$\pm$0.009 & 1.26 \\
A220 & 2.22$\pm$0.02 & =N(H) & 3.1$\pm$0.3 & =N(Al) & =N(Al) & =N(Al) & 1.90$\pm$0.04 & 3.5$^{+0.1}_{-0.2}$ & 0.35 & 100 & 0.121$\pm$0.002 & 0.243$\pm$0.005 & 1.11 \\
A221 & 2.21$\pm$0.02 & =N(H) & 2.6$\pm$0.4 & =N(Al) & =N(Al) & =N(Al) & 1.85$\pm$0.05 & 3.5$^{+0.1}_{-0.2}$ & 0.35 & 100 & 0.119$\pm$0.002 & 0.230$\pm$0.005 & 1.28 \\
A222 & 2.23$\pm$0.02 & =N(H) & 2.7$\pm$0.4 & =N(Al) & =N(Al) & =N(Al) & 1.91$\pm$0.05 & 3.5$^{+0.1}_{-0.2}$ & 0.35 & 100 & 0.120$\pm$0.002 & 0.239$\pm$0.004 & 1.18 \\
\midrule
\multicolumn{3}{l}{\textbf{Rebrightening phase}} \\
\midrule
\nicer\/ & N(H) & N(Mg) & N(Al) & N(Si) & N(S) & N(Ca) & $\Gamma$ & log $\xi$ & R$_{\mathrm{f}}$ & R$_{\mathrm{in}}$ & F$_{\mathrm{abs}}$ & F$_{\mathrm{unabs}}$ & $\chi^{2}_{\mathrm{red}}$ \\
obs. & \multicolumn{6}{c}{10$^{22}$ atoms/cm$^{2}$} & & & & & 10$^{-8}$ erg/s/cm$^{2}$ & 10$^{-8}$ erg/s/cm$^{2}$ & \\
\midrule
B101 & 2.19$\pm$0.01 & 2.9$\pm$0.4 & 9$\pm$2 & 3.4$\pm$0.2 & 4.4$\pm$0.4 & 14$\pm$2 & 2.14$\pm$0.02 & 3.66$^{+0.07}_{-0.05}$ & 0.35 & 100 & 0.151$\pm$0.003 & 0.363$\pm$0.007 & 1.13 \\
B201 & 2.15$\pm$0.02 & 3.5$\pm$0.4 & 7$\pm$3 & 3.4$\pm$0.2 & 4.4$^{+0.6}_{-0.5}$ & 14$\pm$2 & 2.07$^{+0.04}_{-0.03}$ & 3.6$\pm$0.1 & 0.35 & 100 & 0.160$\pm$0.004 & 0.367$\pm$0.008 & 1.15 \\
B301 & 2.19$^{+0.01}_{-0.02}$ & 3.1$\pm$0.4 & 11$\pm$2 & 3.4$\pm$0.2 & 4.8$\pm$0.4 & 14$\pm$2 & 2.16$\pm$0.02 & 3.67$^{+0.07}_{-0.05}$ & 0.35 & 100 & 0.172$\pm$0.004 & 0.422$\pm$0.009 & 1.42 \\
B401 & 2.19$\pm$0.01 & 3.7$\pm$0.4 & 9$^{+2}_{-3}$ & 3.5$^{+0.1}_{-0.3}$ & 4.5$^{+0.3}_{-0.6}$ & 14$\pm$2 & 2.20$^{+0.01}_{-0.04}$ & 3.77$^{+0.05}_{-0.09}$ & 0.35 & 100 & 0.165$\pm$0.004 & 0.419$\pm$0.008 & 1.14 \\
B501 & 2.15$\pm$0.03 & 4.1$\pm$0.9 & 16$^{+5}_{-6}$ & 3.2$\pm$0.4 & 6$\pm$1 & 19$^{+4}_{-5}$ & 2.15$\pm$0.06 & 3.7$^{+0.2}_{-0.1}$ & 0.35 & 100 & 0.148$\pm$0.008 & 0.370$\pm$0.017 & 1.01 \\
B601 & 2.21$\pm$0.01 & 3.5$\pm$0.4 & 11$\pm$2 & 3.7$\pm$0.2 & 5.0$\pm$0.4 & 13$\pm$2 & 2.18$\pm$0.02 & 3.69$^{+0.07}_{-0.03}$ & 0.35 & 100 & 0.177$\pm$0.004 & 0.445$\pm$0.009 & 1.56 \\
B701 & 2.21$^{+0.02}_{-0.01}$ & 3.6$\pm$0.4 & 14$\pm$2 & 3.6$\pm$0.1 & 5.2$\pm$0.3 & 14$\pm$2 & 2.20$\pm$0.01 & 3.71$^{+0.03}_{-0.05}$ & 0.35 & 100 & 0.164$\pm$0.003 & 0.423$\pm$0.008 & 1.21 \\
B801 & 2.22$\pm$0.02 & 4.1$\pm$0.4 & 16$\pm$2 & 4.1$\pm$0.2 & 6.1$\pm$0.4 & 10$\pm$2 & 2.09$\pm$0.02 & 3.70$^{+0.08}_{-0.02}$ & 10$^{+0}_{-6.23}$ & 100 & 0.141$\pm$0.005 & 0.35$\pm$0.01 & 1.31 \\
B901 & 2.29$\pm$0.02 & 4.2$^{+0.5}_{-0.4}$ & 23$\pm$3 & 3.9$\pm$0.2 & 5.4$^{+0.4}_{-0.3}$ & 8$\pm$2 & 2.20$\pm$0.01 & 3.70$\pm$0.02 & 10$^{+0}_{-4.96}$ & 100 & 0.120$\pm$0.002 & 0.327$\pm$0.006 & 1.28 \\
\midrule
\multicolumn{3}{l}{\textbf{Obscured phase}} \\
\midrule
\nicer\/ & N(H) & N(Mg) & N(Al) & N(Si) & N(S) & N(Ca) & $\Gamma$ & log $\xi$ & R$_{\mathrm{f}}$ & R$_{\mathrm{in}}$ & F$_{\mathrm{abs}}$ & F$_{\mathrm{unabs}}$ & $\chi^{2}_{\mathrm{red}}$ \\
obs. & \multicolumn{6}{c}{10$^{22}$ atoms/cm$^{2}$} & & & & & 10$^{-8}$ erg/s/cm$^{2}$ & 10$^{-8}$ erg/s/cm$^{2}$ & \\
\midrule
C001 & 2.21$\pm$0.02 & 4.7$\pm$0.6 & 22$\pm$4 & 3.2$\pm$0.3 & 4.2$\pm$0.8 & 6$\pm$2 & 1.89$^{+0.03}_{-0.06}$ & 3.16$^{+0.03}_{-0.05}$ & $-$2 & 1 & 0.072$\pm$0.002 & 0.161$\pm$0.005 & 1.49 \\
C202 & 2.03$\pm$0.03 & 4.1$^{+0.9}_{-0.8}$ & 27$^{+6}_{5}$ & 3.7$\pm$0.3 & 5.9$\pm$0.9 & =N(H) & 1.44$\pm$0.03 & 2.80$^{+0.03}_{-0.05}$ & $-$2 & 1 & 0.021$\pm$0.002 & 0.031$\pm$0.003 & 1.09 \\
C203 & 2.08$^{+0.05}_{-0.07}$ & 8$\pm$2 & 37$^{+10}_{9}$ & 5.5$^{+0.7}_{-0.6}$ & 10$\pm$2 & =N(H) & 1.64$^{+0.03}_{-0.04}$ & 2.44$^{+0.04}_{-0.06}$ &  $-$2 & 1 & 0.034$\pm$0.002 & 0.055$\pm$0.004 & 1.65 \\
C204 & 2.23$^{+0.03}_{-0.04}$ & 5$\pm$1 & 30$\pm$5 & 3.1$\pm$0.3 & 5.0$\pm$0.8 & =N(H) & 1.69$^{+0.03}_{-0.04}$ & 2.74$\pm$0.01 & $-$2 & 1 & 0.048$\pm$0.003 & 0.084$\pm$0.005 & 1.57 \\
C205 & 2.18$^{+0.04}_{-0.05}$ & 4$\pm$1 & 16$\pm$7 & 2.2$\pm$0.4 & 2$\pm$1 & =N(H) & 1.53$\pm$0.06 & 2.80$^{+0.03}_{-0.05}$ &  $-$2 & 1 & 0.039$\pm$0.004 & 0.059$\pm$0.007 & 1.69 \\
C206 & 2.13$\pm$0.05 & 7$^{+2}_{-1}$ & 29$^{+10}_{9}$ & 3.4$^{+0.6}_{-0.5}$ & 6$\pm$2 & =N(H) & 1.64$^{+0.05}_{-0.06}$ & 2.73$\pm$0.02 & $-$2 & 1 & 0.032$\pm$0.003 & 0.053$\pm$0.006 & 1.47 \\
C207 & 2.09$^{+0.04}_{-0.06}$ & 3$\pm$1 & 25$^{+7}_{8}$ & 1.7$^{+0.4}_{-0.6}$ & =N(H) & =N(H) & 1.51$^{+0.05}_{-0.08}$ & 2.77$^{+0.03}_{-0.04}$ & $-$2 & 1 & 0.016$\pm$0.002 & 0.024$\pm$0.003 & 1.54 \\
C208 & 2.0$\pm$0.1 & 2$\pm$2 & 27$\pm$10 & 2.6$\pm$0.7 & =N(H) & =N(H) & 1.20$^{+0.04}_{-0.01}$ & 2.82$^{+0.02}_{-0.03}$ & $-$2 & 1 & 0.029$\pm$0.001 & 0.036$\pm$0.002 & 1.42 \\
C209 & 2.13$^{+0.1}_{-0.3}$ & 5$^{+5}_{-4}$ & 14$^{+14}_{24}$ & 3$\pm$2 & =N(H) & =N(H) & 1.63$\pm$0.2 & 3.01$^{+0.2}_{-0.1}$ & $-$2 & 1 & 0.011$\pm$0.003 & 0.019$\pm$0.007 & 1.28 \\
C301 & 2.25$\pm$0.04 & 5$\pm$1 & 28$\pm$5 & 2.5$\pm$0.3 & =N(H) & =N(H) & 1.63$^{+0.02}_{-0.03}$ & 2.82$^{+0.02}_{-0.03}$ & $-$2 & 1 & 0.022$\pm$0.001 & 0.037$\pm$0.002 & 1.55 \\
C302 & 1.8$\pm$0.1 & =N(H) & =N(H) & =N(H) & =N(H) & =N(H) & 1.20$^{+0.06}_{-0.01}$ & 3.01$\pm$0.01 & $-$2 & 1 & 0.0062$\pm$0.0005 & 0.0078$\pm$0.0006 & 1.45 \\
C303 & 2.0$\pm$0.1 & =N(H) & 19$^{+14}_{18}$ & =N(H) & =N(H) & =N(H) & 1.3$\pm$0.1 & 2.75$^{+0.02}_{-0.03}$ & $-$2 & 1 & 0.010$\pm$0.002 & 0.012$\pm$0.002 & 1.16 \\
C304 & 1.8$^{+0.1}_{-0.2}$ & =N(H) & =N(H) & =N(H) & =N(H) & =N(H) & 1.2$^{+0.2}_{-0.1}$ & 2.75$^{+0.04}_{-0.02}$ & $-$2 & 1 & 0.009$\pm$0.002 & 0.011$\pm$0.003 & 1.33 \\
\bottomrule
\end{longtable}
}
}

\end{appendix}

\end{document}